\documentclass[10pt, aps, pra, superscriptaddress, nofootinbib, 
               amsmath, 
               twocolumn, 
               showpacs, 
               raggedbottom, 
               floatfix, 
               longbibliography,
               colorlinks=true, linkcolor=blue, citecolor=blue, urlcolor=blue,
               ]{revtex4-2}

\providecommand{\exclude}[1]{}

\newif\ifsubmit
\submitfalse
\submittrue

\ifsubmit
  \usepackage[acronyms,shortcuts,notranslate]{glossaries}
  \newacronym{BEC}{BEC}{Bose-Einstein condensate}
  \newacronym[longplural={charge-coupled devices}]{CCD}{CCD}{charge-coupled device}
  \newacronym{WKB}{WKB}{Wentzel–Kramers–Brillouin}
  \newacronym{mw}{mw}{microwave}
  \newacronym{TF}{TF}{Thomas-Fermi}
  \newacronym{NSF}{NSF}{National Science Foundation}

  \usepackage[breaklinks]{hyperref}
  \usepackage[nameinlink]{cleveref}

  \usepackage{siunitx}
  \usepackage[version=3]{mhchem}
  \mhchemoptions{textfontcommand=\fontencoding{U}\fontfamily{zeur}\selectfont}

  \makeatletter
  \apptocmd{\frontmatter@abstractheading}{\textsf{\textbf{\abstractname}}\newline}{}{}

  \def\bibsection{%
    \expandafter\section\expandafter*\expandafter{\refname}%
    \@nobreaktrue%
  }
  \makeatother

  \usepackage[tracking]{microtype}
  \SetTracking{ encoding = *, shape = sc }{ 25 }

  \usepackage[sc,osf]{mathpazo}
  \usepackage[euler-digits,small]{eulervm}
  \renewcommand{\hbar}{\hslash}
  
  \crefname{table}{Table}{Tables}
  \crefname{figure}{Fig.}{Figs.}
  \crefname{equation}{Eq.}{Eqs.}
  \Crefname{table}{Table}{Tables}
  \Crefname{figure}{Fig.}{Figs.}
  \Crefname{equation}{Eq.}{Eqs.}

  \usepackage{titlesec}
  \titleformat{\section}[hang]
              {\sffamily\bfseries}
              {}
              {0pt}
              {}
  \titlespacing*{\section}       
                {0pt}            
                {*3}             
                {*0}             
                [0pt]

  \titleformat{\subsection}[runin]
              {\bfseries}
              {}
              {0pt}            
              {}[.]
  \titlespacing{\subsection}
               {0pt}           
               {*3}            
               {*1}            
               [10pt]          

  \renewcommand{\figurename}{Fig.}
  \renewcommand{\tablename}{Tab.}

  \makeatletter
  \renewcommand{\fnum@figure}{\textsf{\textbf{\figurename~\thefigure}}}
  \renewcommand{\fnum@table}{\textsc{\textbf{\tablename~\thetable}}}
  \renewcommand{\@caption@fignum@sep}{:\hspace{0.5em}\ignorespaces}%
  \makeatother
  
  \appto\mycaptionhook{%
    
    \sffamily%
  }
  \appto\mysupplementhook{%
    \clearpage\newpage
    \setcounter{page}{1}
    
    \setcounter{figure}{0}
    \renewcommand{\figurename}{\gobble}
    \renewcommand{\thefigure}{Supplementary Fig.~\arabic{figure}}
  }

  \creflabelformat{section}{#2#1#3}  
\else
  \usepackage[nature]{my_paper}
  \usepackage{my_acronyms}
\fi

\usepackage[inline]{enumitem}
\usepackage{eurosym}
\usepackage{mathrsfs}
\usepackage{xspace}
\usepackage{multirow}
\usepackage{tabularx}
\usepackage[normalem]{ulem}
\usepackage{xcolor}
\usepackage{amsfonts}
\usepackage{amsmath}
\usepackage{amssymb}
\usepackage{graphicx}
\usepackage{dcolumn}
\usepackage{bm}

\usepackage[caption=false]{subfig}
\newcommand{\labelphantom}[1]{\subfloat{\label{#1}}}

\graphicspath{{figures/}}

\newcommand{\abs}[1]{\ensuremath{\lvert{#1}\rvert}}
\newcommand{\ket}[1]{\ensuremath{\lvert{#1}\rangle}}

\newcommand{\mat}[1]{\mathbf{#1}}
\newcommand{\pdiff}[2]{\frac{\partial #1}{\partial #2}}
\newcommand{\vect}[1]{\boldsymbol{#1}}

\DeclareSymbolFont{greekletters}{OML}{zplm}{m}{it}
\DeclareMathSymbol{\varsigma}{\mathalpha}{greekletters}{"26}

\begin{document}

\title{Gravitational caustics in an atom laser}

\author{M.~E.~Mossman}
\affiliation{Department of Physics and Astronomy, Washington State University, Pullman, WA, USA 99164}
\affiliation{Department of Physics and Biophysics, University of San Diego, San Diego, CA, USA 92110}

\author{T.~M.~Bersano}
\affiliation{Department of Physics and Astronomy, Washington State University, Pullman, WA, USA 99164}

\author{Michael McNeil Forbes}
\email{m.forbes@wsu.edu}
\affiliation{Department of Physics and Astronomy, Washington State University, Pullman, WA, USA 99164}
\affiliation{Department of Physics, University of Washington, Seattle, WA, USA 98105}

\author{P.~Engels}
\email{engels@wsu.edu}
\affiliation{Department of Physics and Astronomy, Washington State University, Pullman, WA, USA 99164}

\begin{abstract}
  Typically discussed in the context of optics, caustics are envelopes of classical trajectories (rays) where the density of states diverges, resulting in pronounced observable features such as bright points, curves, and extended networks of patterns.
  Here, we generate caustics in the matter waves of an atom laser, providing a striking experimental example of catastrophe theory applied to atom optics in an accelerated (gravitational) reference frame.
  We showcase caustics formed by individual attractive and repulsive potentials, and present an example of a network generated by multiple potentials.
  Exploiting internal atomic states, we demonstrate fluid-flow tracing as another tool of this flexible experimental platform.
  The effective gravity experienced by the atoms can be tuned with magnetic gradients, forming caustics analogous to those produced by gravitational lensing.
  From a more applied point of view, atom optics affords perspectives for metrology, atom interferometry, and nanofabrication.
  Caustics in this context may lead to quantum innovations as they are an inherently robust way of manipulating matter waves.
\end{abstract}

\maketitle
\clearpage

\section{Introduction}
From light refracted by a sheet of glass to the light patterns seen on the ocean floor, rainbows, or the observation of gravitational lensing, caustics play a central role in the way optics presents itself in nature~\cite{Berry1980,Nye1999}.
Unlike foci produced by optical instruments, caustics are generic in the sense that they do not need very specialized circumstances to exist, and are structurally stable~\cite{Nye1999}, leading to their widespread occurrence.
They are formed, for example, when light is reflected (catacaustic) or refracted (diacaustic) from a curved surface~\cite{Berry1980,Nye1999,Born1999,Weinstein1969}.
While most visible in optics -- prototypical caustics can readily be observed with polarized, coherent light~\cite{Berry1980,Berry1981,Marston1984,Kaduchak1994,Borghi2016} -- the phenomenon of caustics and the underlying catastrophe theory~\cite{Thom:1975, Arnold:1986} have found far reaching interest.
For example, caustics and catastrophe theory have been discussed in the context of generic two-mode quantum systems~\cite{ODell:2012,Mumford:2017,Mumford:2019}, nuclear physics~\cite{DaSilveira:1973}, general relativity~\cite{Hasse_1996}, social sciences~\cite{Oliva:1981}, and robotics~\cite{Carricato:2002}. 
Caustics have also come into the focus of studies with electron microscopes where they may be exploited for advanced imaging techniques~\cite{Petersen:2013}.

Ultracold quantum gases provide a flexible platform for performing atom-optics experiments~\cite{Shimizu2002,Balykin2003,Oberst2005,Kouznetsov2006,Rohwedder2007,Cronin2009,Dall:2010,Simula2013} where cold atoms, instead of photons, are used to generate atom-optical components with possible applications for fundamental science, atom interferometry, metrology, and new nano-fabrication approaches.
Catastrophe atom optics -- the formation of caustics by atom trajectories -- has previously been discussed in specific settings, for example in the context of atoms being released from a magneto-optical trap~\cite{Rooijakkers:2003,Rosenblum2014}, atoms diffracting from a one-dimensional optical lattice~\cite{ODell:2001,Huckans:2009}, or expanding \glspl{BEC} with spatially varying initial phase~\cite{Chalker:2009}.
To study caustics in matter waves, a particularly powerful tool and natural setting is an atom laser~\cite{Mewes1997, Naraschewski1997, Ketterle:1997, Steck1998,Bloch1999, Schneider1999, Ballagh:2000, Bloch:2000, LeCoq2001, Bloch:2001, Chikkatur2002, Haine:2002, Lee:2015, Harvie:2020, Riou:2008}, which is a coherent stream of collimated atoms out-coupled from a dilute-gas \gls{BEC}.
While caustics and catastrophe theory have been used to characterize the atom laser itself, in particular in terms of its transverse beam profile~\cite{Busch:2002, Kohl:2005, Riou:2006, Dall:2007}, here we exploit the atom laser as a source of flow interacting with external potentials to generate a broad variety of caustic features.
These features include individual fold and cusp caustics, and even complex caustic networks.

\begin{figure}[tbp]
  \centering
  \labelphantom{fig:fig1a}
  \labelphantom{fig:fig1b}
  \labelphantom{fig:fig1c}
  \includegraphics[width=\columnwidth]{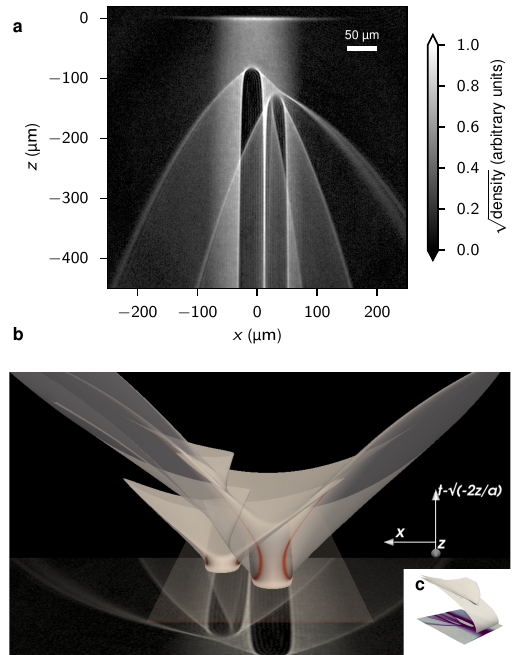}
  \caption{Caustic networks from two repulsive potentials in an atom laser.
    \textbf{a} Experimental data averaged over 15 runs. The atom laser propagates from top ($z=0$) to bottom.
    Each barrier is Gaussian-shaped [\cref{eq:U}] with $U_0/k_B=\SI{23.1}{µK}$ and $\sigma=\SI{10.3}{µm}$.
    The left barrier is positioned $h=\SI{85}{µm}$ below the injection point of the atoms ($\varepsilon = \num{0.95}$, see \cref{eq:epsilon}).
    The right barrier is offset from this by $(\Delta x, \Delta z) = (39,-42)~\mu$m ($\varepsilon = \num{0.64}$).
    \textbf{b} A numerical rendering of the classical trajectories as a sheet $(x_i, t_i) \mapsto (x, z, t_i - \sqrt{-2z/a_z})$, the singularities of which appear as caustics when projected down into the $(x, z)$ imaging plane where the experimental data is shown again.
    Here, $t$ is the vertical direction, while $x$ increases to the left, and $z$ decreases into the image.
    Red shading denotes regions where $\det{\mat{J}}^{-1}$ is large [\cref{eq:detJ}], corresponding to the caustics when projected onto the imaging plane.
    \textbf{c} A numerical rendering of the classical trajectories as a sheet $(x_i, t_i) \mapsto (x, z, t_i)$, which now includes the free-fall background.
    }
  \label{fig:fig1}
\end{figure}

On large scales, the networks formed by caustics can be quite intricate.
An example is shown in \cref{fig:fig1}, where two repulsive optical Gaussian potentials are placed in the atom laser beam.
Caustics arise from singularities in the continuous map $(x_i, t_i) \mapsto (x, z)$ of atoms injected at time $t=-t_i$ and position $(x=x_i, z=0)$ to the imaging plane $(x, z)$ at the time of imaging $t=0$.
As shown in \cref{fig:fig1b}, we can visualize this as a sheet embedded in three-dimensions $(x, z, t)$: the caustics occur where this sheet has vertical tangents, colored red in the figure.
Despite the intricacy of the produced patterns, catastrophe theory reveals that all generic features can be categorized.
In this case, we observe two stable types of singularities -- folds and cusps.

To explore and classify the generated flow patterns, we apply a variety of flow-visualization techniques, including experimental fluid-flow tracing based on internal-state manipulation of the atoms, and simulated three-dimensional folded sheets.
We achieve quantitative agreement between theory and experimental results.
Atom optics differs from terrestrial light optics in a variety of ways, including the ability to easily introduce attractive as well as repulsive potentials, the power to perform internal state manipulation, e.g., for fluid flow tracing, and the ability to study non-negligible effects of gravity.
The accelerated reference frame in our experiments, which is due to the presence of an effective gravity that can be adjusted using magnetic gradients, is natural for atom optics, and leads to features that would not exist otherwise, such as the transition from attached fan-like features to a detached crescent-shape caustic in the case of a repulsive potential.
In contrast to previous work, the use of an atom laser combined with external potentials and an adjustable effective gravity provides a general platform for catastrophe atom optics investigations that can lead to rich observable dynamics with applications to cosmological effects.
The direct imaging of sharply delineated features, a wealth of observable phenomena dependent on the shape, strength and sign of the potential(s), and the ability to trace atomic flow make this setting a very powerful platform.

\begin{figure}
  \centering
  \includegraphics{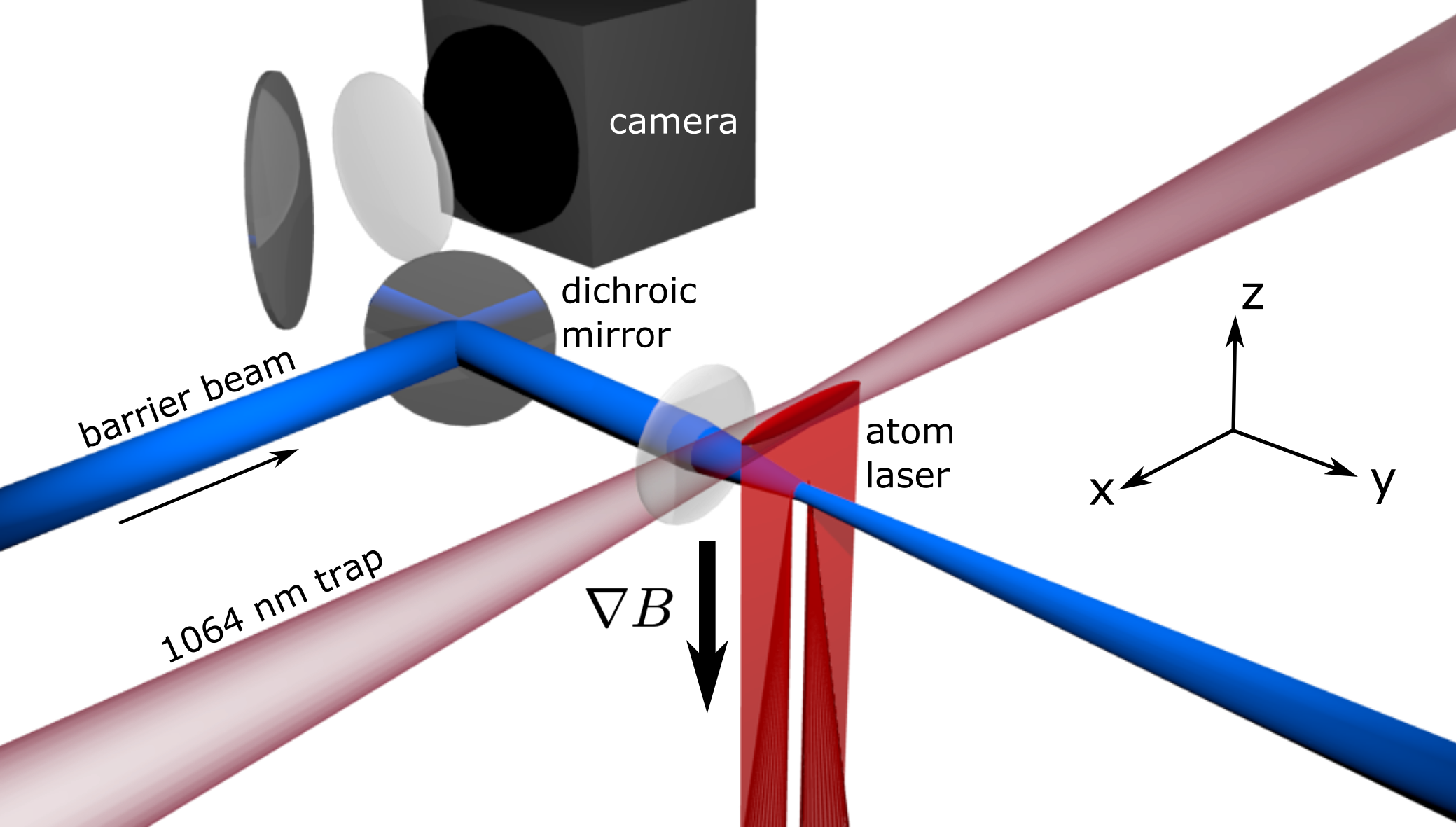}
  \caption{Experimental setup.  Atoms from a \ce{Rb} \gls{BEC} (red) are coherently out-coupled from a dipole trap (maroon) to form an atom laser.
    The atoms are accelerated downward by the combined action of gravity and an external magnetic gradient.
    The accelerated flow encounters a potential created by an additional laser (blue).
    In this depiction, the potential is repulsive to the atoms.
    The flow profile is imaged by a \gls{CCD} camera using absorption imaging.
  }
  \label{fig:setup}
\end{figure}

\section{Results}
\subsection{Experimental Setup}
The experimental setup for these investigations is schematically depicted in \cref{fig:setup}.
A \ce{^{87}Rb} \gls{BEC} is initially confined in an elongated harmonic trap.
The trap is formed by a combination of an attractive focused dipole laser and an additional magnetic field gradient (see Methods section for details).
In this setup, the $x$-axis is oriented along the weakly confined direction of the trap, $y$ along the imaging axis, and $z$ vertically.
The \gls{BEC} is prepared in the $\ket{F, m_F} = \ket{1,-1}$ spin state, which is supported against gravity by the trap.
A \SI{7}{ms} long \gls{mw} tone is used to coherently transfer atoms to the $\ket{F, m_F} = \ket{2,-2}$ spin state, which is magnetically expelled out of the trap in the downward direction with an acceleration of $a_z = \SI{26.9(3)}{µm/ms^2} \approx 2.75g$.   
This accelerated, collimated stream of atoms forms the atom laser. 
A laser propagating in the positive $y$-direction generates an attractive or repulsive potential.
Imaging is performed along the negative $y$-direction, and a dichroic mirror is used to overlay the potential beam path with the imaging laser.

We note that our observation geometry differs from that typically used in optics where a wavefront passes through a refracting surface and then propagates to a screen on which it is observed.
In our case, the images include the propagation direction as one axis of the two-dimensional observation plane.

\subsection{Folds and cusps}
To prepare for the discussion of our experimental observations, we begin with a brief theoretical review of caustics. 

As we shall show below, the prominent features in our experiment can be well described by analyzing the classical dynamics of the atoms. 
In classical mechanics (optics), caustics occur on the envelope of classical trajectories $q(t)$ where the trajectories (rays) and their tangents converge.
These classical trajectories are stationary points of the action $S'[q] = 0$, and the caustics arise when the hessian $S''(q)$ is degenerate in some directions.
These overlapping trajectories thus lead to an infinite density of states -- classical divergences apparent to anyone who has accidentally started a fire from a curved mirror -- that are softened by quantum mechanics where they are characterized by a breakdown in the \gls{WKB} approximation~\cite{DeWitt-Morette:1983, Cartier:2006, Mumford:2019}.

The atom laser continuously injects a thin line of atoms at height $z_i=0$, which then fall under a constant acceleration $a_z$ and are scattered by the potential.
The injection region has limited spatial extent ($\sim \SI{1}{µm}$ in the $y$ and $z$ directions), and we assume $z_i = y_i \approx 0$ in our analysis.

Geometrically, this input can be described by a uniform sheet in the two-dimensional \textit{state space} $(x_i, t_i)$ spanned by the initial injection sites of the atom laser $x=x_i$, $z=0$, at  time $t=-t_i$.
The atoms then follow classical trajectories, falling under the acceleration $a_z$, scattering from the optical potential(s), and ending at a final location $(x, z)$ at the time of imaging, which we take as $t=0$ so that $t_i$ has the interpretation of the time over which the atoms have fallen.
Assuming a thin initial stream of atoms, the images represent a continuous mapping of $(x_i, t_i)$ state space into the $(x, z)$ imaging plane through an intermediate sheet $(x, z, t_i - \sqrt{-2a_z/z})$ which we show in \cref{fig:fig1b}.
This representation removes the effect of the background acceleration $a_z$ from the visualization and accentuates the observed features.

Caustics correspond to singularities in this mapping.
Assuming a constant injection rate -- a good approximation for our experiments -- the observed local density is inversely proportional to the determinant of the Jacobian of the mapping
\begin{gather}\label{eq:detJ}
  \det(\mat{J}) = \begin{vmatrix}
    \pdiff{z}{x_i} & \pdiff{z}{t_i}\\
    \pdiff{x}{x_i} & \pdiff{x}{t_i}
  \end{vmatrix},
\end{gather}
which becomes zero at the points corresponding to the caustics.
These classical divergences are softened by quantum mechanics~\cite{DeWitt-Morette:1983, Cartier:2006}, resulting in an Airy-function interference pattern.
The size of these features $\sqrt[3]{4\hbar^2/m^2a_z} \sim \SI{0.5}{µm}$ is too small to observe in the current experiment, but ripe for future study.

Whitney~\cite{Whitney:1955} proved that continuous mappings from a plane into a plane can result in only two stable types of singularity -- folds and cusps, both of which are observed here.
Folds appear as curves where the surface in our three-dimensional embedding $(x, z, t)$ has vertical tangents along $t$, and cusps appear as singular points where these folds meet.
These are the only singularities that are stable, in the sense that they will persist, for example, even if the imaging angle is changed slightly.
Additional non-generic singularities can in principle be seen with this geometry, but these must be artificially tuned, for example by intentionally aligning cusp caustics.
Arnold~\cite{Arnold:1986} provides a complete classification of these singularities.

We demonstrate here both fold and cusp caustics by inserting a Gaussian optical potential at position $z = -h$:
\begin{gather}\label{eq:U}
  U(x, z) = U_0\exp\left(-\frac{x^2 + (z+h)^2}{\sigma^2/2}\right),
\end{gather}
where $\sigma$ is the Gaussian waist and $U_0$ is the central strength of the potential.
After scattering through this potential, classical particles will eventually move on parabolic trajectories.
Qualitatively, the nature of the scattering and the associated caustics will be largely governed by the ratio
\begin{gather}
  \label{eq:epsilon}
  \varepsilon = \frac{U_0}{ma_z h}.
\end{gather}
The caustic structure changes dramatically at $\varepsilon \approx 1$.
In the experiments, the sign and magnitude of $\varepsilon$ can be varied over a wide range, for example by adjusting the wavelength and intensity of the laser generating the optical potential.

\begin{figure*}
    \centering
    \includegraphics{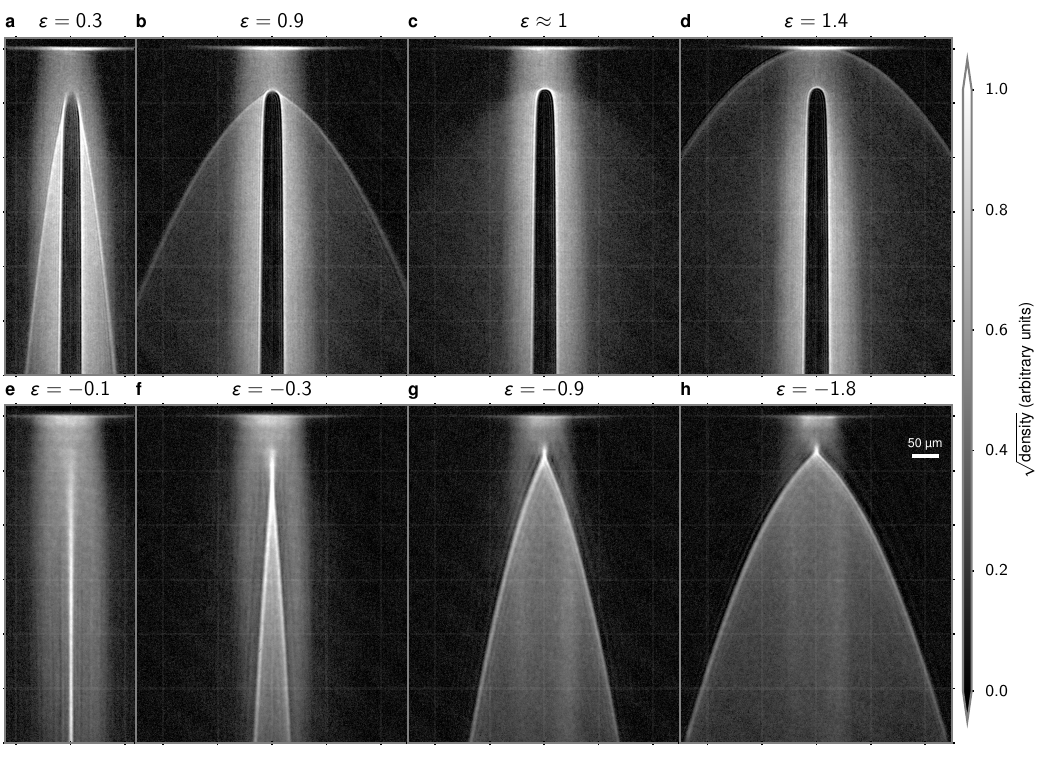}
    \labelphantom{fig:GaussShock-r1}
    \labelphantom{fig:GaussShock-r2}
    \labelphantom{fig:GaussShock-r3}
    \labelphantom{fig:GaussShock-r4}
    \labelphantom{fig:GaussShock-a1}
    \labelphantom{fig:GaussShock-a2}
    \labelphantom{fig:GaussShock-a3}
    \labelphantom{fig:GaussShock-a4}
    \caption{Experimental observation of flow with repulsive or attractive potentials.
      \textbf{a}-\textbf{d} A repulsive potential is centered $h=\SI{78(1)}{µm}$ below the trapped \gls{BEC}.
      \textbf{e}-\textbf{h} An attractive potential is centered $h=\SI{46(1)}{µm}$ below the trapped \gls{BEC}.
      Individual panels are labeled by the energy ratio $\epsilon$ as defined in the main text. 
      All images have been averaged over 6 independent experimental images with the same parameters.
      A faint grid is overlaid at \SI{100}{µm} increments.
      The scale for square-root of the density has been set so that pure white corresponds to the
      99.95\textsuperscript{th} percentile over all images to emphasize the caustic structure.
      The slight fringing seen in the images is due to optical effects in the imaging system, not matter-wave interference.
    }
    \label{fig:GaussShock}
\end{figure*}

Typical results are shown in \cref{fig:GaussShock}.
The top row (\Crefrange{fig:GaussShock-r1}{fig:GaussShock-r4}) was generated by inserting a \emph{repulsive} potential ($\varepsilon > 0$) in the atom laser, while the bottom row (\Crefrange{fig:GaussShock-a1}{fig:GaussShock-a4}) was generated by inserting an \emph{attractive} potential ($\varepsilon < 0$).

\begin{figure*}
    \centering
    \includegraphics{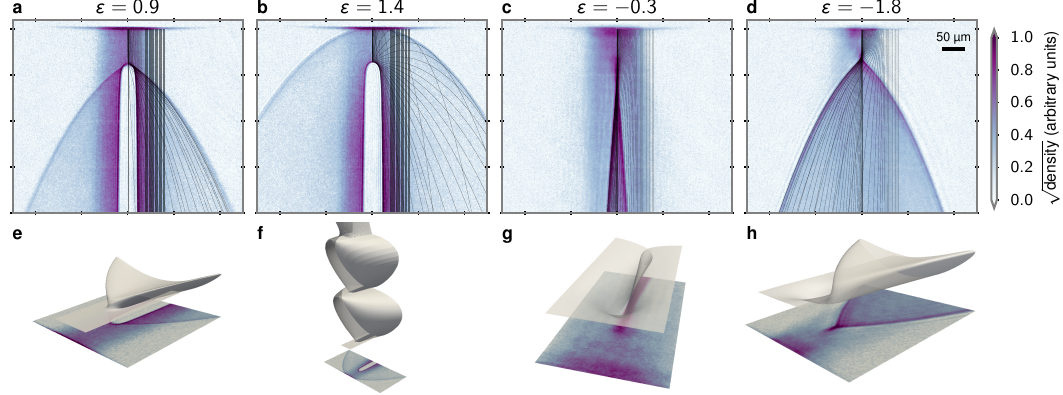}
    \labelphantom{fig:GaussShockTh-r2}
    \labelphantom{fig:GaussShockTh-r4}
    \labelphantom{fig:GaussShockTh-a2}
    \labelphantom{fig:GaussShockTh-a4}
    \labelphantom{fig:GaussShockTh-r2-3d}
    \labelphantom{fig:GaussShockTh-r4-3d}
    \labelphantom{fig:GaussShockTh-a2-3d}
    \labelphantom{fig:GaussShockTh-a4-3d}
    \caption{Numerical simulation and visualization.
    \textbf{a-d} Numerical results of \num{50} trajectories from $x_i \in $~\SIrange{0}{80}{µm} uniformly spaced in $x_i^{1/2}$ to emphasize the scattering, overlaid on the experimental data.
    For clarity, a reversed color scheme has been used to plot the experimental images. 
    \textbf{e-h} Corresponding map $(x_i, t_i)\mapsto (x, z, t_i-\sqrt{-2z/a_z})$ with the vertical time axis scaled by a factor of 5 for frame \textbf{g}.
    These numerics are generated for a single repulsive (\textbf{a}, \textbf{b}, \textbf{e}, \textbf{f}) or attractive (\textbf{c}, \textbf{d}, \textbf{g}, \textbf{h}) potential with the energy ratio noted at the top of each panel.
    }
\label{fig:GaussShockTh}
\end{figure*}

\subsection{Repulsive potential}
The results presented in \crefrange{fig:GaussShock-r1}{fig:GaussShock-r4} have been obtained using a repulsive Gaussian potential. 
The potentials were generated by a laser with a wavelength of \SI{660}{nm} and a beam waist of $\sigma \approx \SI{11.3(5)}{µm}$ located $h \approx \SI{78(3)}{µm}$ below the trapped \gls{BEC}.
The ratio $\varepsilon$ was varied by changing the laser intensity. 
For low values of $\varepsilon$, pronounced fan-like features are seen to emanate from both sides of the repulsive potential (\cref{fig:GaussShock-r1}).
The observed edge steepness of these features appears to be limited only by our imaging resolution of approximately \SI{3}{µm}.
Our analysis presented below in the context of \cref{fig:GaussShockTh} identifies these edges as fold caustics.
As the strength of the repulsive potential is increased, the attached fan-like features increase in width (\cref{fig:GaussShock-r2}), ultimately detaching from the potential at $\varepsilon\approx1$ (\cref{fig:GaussShock-r3}) and forming a half-ring-shaped detached caustic as $\varepsilon$ increases above unity (\cref{fig:GaussShock-r4}).
This crossover occurs at a point where, at zero impact parameter, a falling classical atom would slow to rest at the top of the potential, signifying the reflection/transmission threshold.

The detached caustic here is a unique feature of atom optics in a sloped potential (such as the one generated by a constant downward acceleration) and would not exist in the absence of such a slope.
Unlike the fan-like feature, the shape of this detached feature does not change as the potential height of the barrier is further increased.
This is consistent with the analysis presented in \nameref{sec:methods} which shows that any dependence should appear only weakly through effects related to the finite size of the potential.
The observed features presented here are independent of the atomic density; the density of the atom laser is sufficiently low that mean-field effects can be neglected.

These features are reminiscent of the shockwaves created by a supersonic object moving through a fluid (see Refs. ~\cite{VanDyke:1982,Samimy:2004} for classic examples).
In particular, for $\varepsilon < 1$, the caustics look like an attached \textit{oblique shock} (\crefrange{fig:GaussShock-r1}{fig:GaussShock-r2}), while for $\varepsilon > 1$, the shape of the pronounced caustic appearing above the potential (\cref{fig:GaussShock-r4}) resembles that of a detached \textit{bow shock}.
The transition between these occurs for $\varepsilon \approx 1$, and is shown in \cref{fig:GaussShock-r3} which shows faint signatures of both types of caustic.
While the features in our experiment can be well-described by classical free-particle dynamics and, unlike shocks, do not involve non-linear self-steepening, this analogy is intriguing.
In our experiments, the atoms scatter from the potential with an impact velocity of about \SI{6.5}{cm/s}.
For comparison, even in the dense region of the trapped \gls{BEC}, the bulk speed of sound $c_\mathrm{s}\approx\SI{3}{mm/s}$ is more than an order of magnitude smaller.
Thus, while our experiments operate in a regime where mean-field effects play no role, one can envision designing a similar procedure to explore supersonic, or even hypersonic, shockwaves~\cite{Chandrasekhar:1943}.
It should be noted that our observations of caustics are distinct from the observation of Bogoliubov-Cherenkov radiation~\cite{Henson:2018}.

To provide a clearer view on the physics behind the observed features, \cref{fig:GaussShockTh-r2,fig:GaussShockTh-r4} show a comparison to numerical simulation for two repulsive potential strengths, along with corresponding visualizations in the form of folded sheets in \cref{fig:GaussShockTh-r2-3d,fig:GaussShockTh-r4-3d}, respectively.
As in \cref{fig:fig1b}, the projection of the sheets onto the imaging plane reveal the caustics: Caustics occur along singularities of the map [\cref{eq:detJ}], which correspond to portions of this surface with vertical slopes. 
The repulsive potential causes the sheet to fold back on itself without intersections, forming multiples of four fold caustics as one descends (\cref{fig:GaussShockTh-r2-3d,fig:GaussShockTh-r4-3d}). 
Cusp caustics occur where the number of fold caustics changes.

For $0 < \varepsilon < 1$, the sheet overlaps at most three times between the fan-like caustics (see \cref{fig:GaussShockTh-r2,fig:GaussShockTh-r2-3d}).
In the absence of an external acceleration, the maximally scattered trajectory always scatters less than \SI{90}{\degree} for Gaussian potentials.
This implies that the outermost fan caustic is not an envelope, but is instead a single parabolic trajectory (see \nameref{sec:methods}).
In the transition region $\varepsilon \approx 1$, the maximal scattering angle rapidly increases from \SI{90}{\degree} to \SI{180}{\degree}. 
For Gaussian potentials, this region exists only because of the finite acceleration $a_z \neq 0$, and for our parameters is limited to $\num{0.997} \lesssim \varepsilon < 1$ (see \nameref{sec:methods}).

As $\varepsilon$ approaches unity, the abrupt change of the dynamics observed in the experiments corresponds to a drastic change in the sheet structure (\cref{fig:GaussShockTh-r4-3d}).
In terms of the classical trajectories, for $\varepsilon \geq 1$, the central particle with zero impact parameter $x_i = 0$ bounces infinitely many times.
This leads to an infinite series of overlapping sheets as shown in \cref{fig:GaussShockTh-r4-3d}.
In the experimental image \cref{fig:GaussShock-r4}, these collapse to a single feature that looks like a detached bow shock.
Quantum mechanics softens this structure since particles can tunnel through the barrier, but on a scale that we cannot resolve in this experiment.

\begin{figure}[tb]
  \centering
    \includegraphics{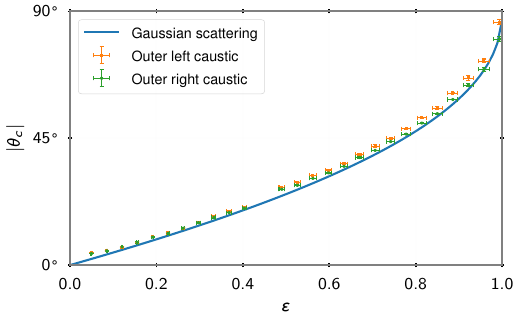}
    \caption{
      Maximal scattering angle $\abs{\theta_c(\varepsilon)}$ vs.\ $\varepsilon$. Angle obtained by fitting \cref{eq:caustic} to the outer left (orange squares) and right (green circles) caustics in the experimental images.
      The standard deviation at each $\epsilon$ provides an estimate of the uncertainty based on three data points.
      The solid line is the classical result $\theta_c(\varepsilon)$ for a particle with kinetic energy $K=mv^2/2 = \varepsilon U_0$ scattering off of a Gaussian potential.
      The small differences between the left and right $\theta_c$ above $\varepsilon \approx 0.5$ is likely due to slight asymmetries in optical potential.
    }
    \label{fig:theta}
\end{figure}

To provide a quantitative analysis, we measure the scattering angle of the caustic $\theta_c(\varepsilon)$ which is dependent on the strength of the potential $\varepsilon$.
As shown in \nameref{sec:methods}, the outer caustics of these fans correspond to a particular parabolic trajectory:
\begin{gather}
  \label{eq:caustic}
  z(x) = z_* - \frac{(x-x_*)^2}{4h \sin^2\theta_c},
\end{gather}
where the maximum of the parabola $(x_*, z_*)$ depends on details of the scattering.
In \cref{fig:theta} we compare $\theta_c(\varepsilon)$ extracted by fitting \cref{eq:caustic} to the outer caustics of the experimental images, with the values computed from the classical scattering problem.
As discussed in \nameref{sec:methods}, the fact that $\abs{\theta_c} < \SI{90}{\degree}$ in this range is a peculiar feature of Gaussian potentials.
The quantitative agreement seen between the experiment and classical scattering allows one to use caustics in the spirit of inverse classical scattering theory. 
Solving the inverse scattering problem provides sensitive input to calibrate properties of the potentials used in an experiment.
For example, slight deviations from the theory, such as an asymmetry between left and right caustics, likely indicate slight asymmetries in the optical potentials.
In this way, catastrophe atom optics can be used as a tool to precisely measure spatial properties of experimental potentials when designing atomtronic devices.
For more information about the analysis of \cref{fig:theta}, see \nameref{sec:methods}.

\subsection{Attractive Potential}
The great flexibility afforded by atom optical techniques allows one to not only change the strength of the potential but also its sign. 
In our experiment, we introduce an attractive potential by using a laser with a wavelength of \SI{850}{nm} and a Gaussian beam waist of $\sigma \approx \SI{27}{µm}$ located $h \approx \SI{46}{µm}$ below the \gls{BEC}.
This attractive potential is an analog of a lens which focuses the atom laser.
For very low potential depths (\cref{fig:GaussShock-a1}), the focusing is weak and the potential appears to produce a nearly collimated, narrow stream of increased density below the potential.
As the potential depth is increased, the parabolic trajectories from either side of the potential are seen to cross (\crefrange{fig:GaussShock-a2}{fig:GaussShock-a4}), and the structure is that of two fold caustics emerging from a cusp (\cref{fig:GaussShockTh-a2-3d,fig:GaussShockTh-a4-3d}).
As with the repulsive potential, for maximal scattering less than \SI{90}{\degree} -- which is the case for $\num{-2.59} \lessapprox \varepsilon < 1$ (see \nameref{sec:methods}) -- these caustics are also parabolic trajectories [\cref{eq:caustic}].
\cref{fig:GaussShockTh-a2,fig:GaussShockTh-a4} show a comparison with classical trajectory calculations with corresponding sheet visualizations in \cref{fig:GaussShockTh-a2-3d,fig:GaussShockTh-a4-3d}. 
Here, the attractive potential draws the sheet downward, causing it to eventually self-intersect. 
This allows for the formation of zero or two fold caustics as one descends.
A central cusp caustic occurs where the number of fold caustics changes from zero to two.
As for the repulsive potentials, excellent quantitative agreement is found between numerics and the experimental results.

\begin{figure}
  \centering
  \includegraphics{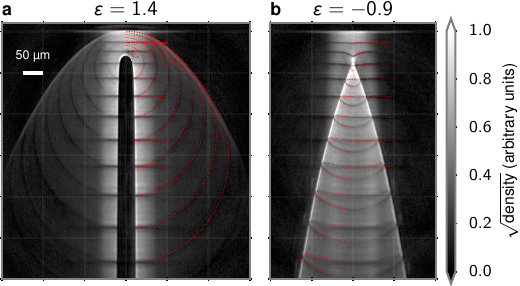}
  \labelphantom{fig:FFtrace-r}
  \labelphantom{fig:FFtrace-a}
  \caption{
    Fluid-flow tracing in the presence of a repulsive or attractive potential.
    \textbf{a} A repulsive potential with waist $\sigma \approx \SI{12}{µm}$ is located $h=\SI{68.0(5)}{µm}$ below the trapped \gls{BEC}. 
    A small horizontal stripe of atoms (located in the region between the trapped \gls{BEC} and the potential) is converted back to the \ket{1,-1} state by a series of brief \SI{100}{µs} \gls{mw} pulses in \SI{0.5}{ms} intervals, leaving a void of atoms in the imaged \ket{2,-2} state.
    Overlaying the experimental image are the corresponding classical trajectories at these fixed $t_i$, shown as dotted (red) curves for positive initial impact parameters $0 < x_0 < \SI{100}{µm}$.
    These correspond to slices through the embedded sheet $(x_i, t_i) \mapsto (x, z, t_i)$ (see for example \cref{fig:fig1c}), demonstrating that these fluid-flow tracers provide a direct way to visualize the folding of this sheet which gives rise to the caustics.
    \textbf{b} Same as (a), but for an attractive potential with waist $\sigma\approx \SI{25}{µm}$ located $h=\SI{50(5)}{µm}$ below the trapped \gls{BEC}.
    Both of the experimental images here have been averaged over 101 runs.
} \label{fig:FFtrace}
\end{figure}

\subsection{Fluid flow tracing}
Going beyond the imaging of the overall fluid flow pattern, internal state manipulation of the atoms affords further powerful ways to visualize and analyze the flow: flow tracers can be created that indicate the evolution of a wavefront as it propagates along the atom laser stream.
This technique is demonstrated in \cref{fig:FFtrace} where a horizontal line of atoms, located in the region between the trapped \gls{BEC} and the potential, is transferred into a state that appears dark in the absorption images.
The transfer is effected by a \SI{100}{µs} \gls{mw} pulse on the $\ket{2,-2}$ to $\ket{1,-1}$ transition.
Spatial selectivity is possible due to the magnetic gradient in which the experiments are performed.
The dark line flows with the atom laser, tracing slices of specific evolution times.
In \cref{fig:FFtrace}, a series of such lines has been injected into the stream at a fixed position between the trapped \gls{BEC} and the potential in \SI{0.5}{ms} time intervals.

Upon scattering from a strong repulsive potential (\cref{fig:FFtrace-r}), full dark rings are observed to propagate away from the potential and connect to the bow-shaped caustic, indicating the wavefront after the scattering event.
Once the particles have fallen far enough, they are no longer influenced by the potential, and these bands are approximately circular arcs of radius $t_i\sqrt{2a_zh}$ centered at height $-h + (a_z-g)t_I^2/2 - a_zt_i^2/2$, extending over $\theta \in [-\theta_c, \theta_c]$ (see \nameref{sec:methods}).
This provides a very visual explanation for the emergence of the caustic as an envelope of rays.

Fluid flow tracing in the case of an attractive potential is shown in \cref{fig:FFtrace-a}. 
The image clearly reveals how the initially horizontal fluid tracer lines are drawn into the region of the potential, from which they emerge as loop structures. 
The left and right apexes of these loops reveal the position of the fold caustics, providing an independent and direct experimental visualization of the formation of the caustics.
Red dotted lines in \cref{fig:FFtrace} show the results of classical trajectories, which are in agreement with the experiment.

\section{Discussion}
In contrast to catastrophe optics of light, catastrophe atom optics has unusual features and provides unique opportunities.
For example, the pronounced half-ring-shaped caustic appearing for sufficiently strong repulsive potentials would not exist without the influence of gravity, and a similar feature would be difficult to see with light outside of cosmological contexts.
The experiments presented here demonstrate the ability to generate and directly image complex caustics in an accelerated reference frame.
These experiments build upon and contribute to the field of catastrophe atom optics by introducing a powerful experimental setting for investigations.

This system suggests an intuitive approach for visualizing the origin of the generic cusp and fold caustics proved by Whitney~\cite{Whitney:1955}.
Real-time dynamics provide a natural embedding of the mapping whose singularities define the caustics, and our technique of fluid-flow tracing generates direct experimental images of slices through this embedding, aiding the interpretation of such atom optics experiments.

The complexity of the observed phenomenology suggests many interesting future extensions of this work, including the construction of more complex caustic networks and the study of departures from classical catastrophe theory as inter-atomic interactions and quantum interference effects start to appear.
Furthermore, our demonstration of fluid-flow tracing might find interesting applications in the study of quantum turbulence or quantum shocks, where following fluid tracers can help to reveal underlying flow patterns.

Our theoretical investigations complement these experiments by using a range of techniques.
The results presented here are based on classical trajectories, providing a concrete demonstration of the principles behind classical catastrophe theory~\cite{Thom:1975, Arnold:1986} in a context beyond the usual setting in optics.
Extending these to include quantum effects can proceed through a semi-classical
expansion, with leading order corrections given by the \gls{WKB} approximation~\cite{DeWitt-Morette:1997, Cartier:2006, Adhikari:2008b,Goussev:2013}, and full quantum effects included numerically (see e.g.~\cite{Edwards:1999, Schneider:1999, Schneider:2000, Mumford:2019}). As a future direction, the excellent agreement with the experiment allows us to validate these different approaches, and thus to design future experiments that will probe higher-order corrections.
This work also lays the foundation to further explore the quantum ramifications of caustics:  Appearing as singularities in the classical action~\cite{DeWitt-Morette:1983, Cartier:2006, Mumford:2019}, caustics play a similar role to quantum scars~\cite{Heller:1984}, giving rise to robust features in the presence of classical chaos.  Atom lasers have a distinct advantage over optics here in that non-linear interactions can also be manipulated.

From a point of view of applications, it is worth noting that caustics also occur in electron microscopes~\cite{Petersen:2013}.  One can imagine an analogous setup for an atom laser, where a specimen under study is placed into the atom laser stream by an optical tweezer operated at the tune-out wavelength~\cite{Arora:2011} for Rb so that the trap itself does not perturb the atom laser. In this context, the experimental platform presented here provides a flexible model system for studying the effects of advanced beam shaping techniques.

\section{Methods}
\label{sec:methods}
\subsection{Experimental setup}
To investigate the dynamics of an atom laser scattered by a Gaussian potential, a \ce{^{87}Rb} Bose-Einstein condensate composed of $5.5\times10^5$ atoms is initially prepared in the $\ket{F, m_F} = \ket{1,-1}$ spin state.
The condensate is confined in a hybrid trap consisting of a single dipole trap with a waist of \SI{20}{µm} and a magnetic quadrupole field which has been vertically shifted above the dipole trap.
This leads to a hybrid trap with trap frequencies of $\{\omega_x, \omega_y, \omega_z\} = 2\pi\times\{7.1, 167, 180\}\;\si{Hz}$.
The initial condensate in the $\ket{F, m_F} = \ket{1,-1}$ spin state has a \gls{TF} radius in the $x$ direction of \SI{93}{µm}.

Atoms are then ejected from the trap by resonantly exciting atoms from the  $\ket{1,-1}$ spin state to the $\ket{2,-2}$ state using a \SI{7}{ms} \gls{mw} pulse.
Atoms in the $\ket{2,-2}$ state accelerate downwards at $a_z = \SI{26.9(3)}{µm/ms^2}$ in the negative $z$-direction. 

The repulsive (attractive) Gaussian potential is created using a \SI{660}{nm} (\SI{850}{nm}) laser focused to a $\sigma=\SI{11.3(5)}{µm}$ (\SI{27(1)}{µm}) Gaussian waist located $h=\SI{78(3)}{µm}$ (\SI{46(3)}{µm}) below the trapped \gls{BEC}.
In the case of \cref{fig:fig1a}, where two repulsive potentials were used, orthogonal polarizations of light were used to prevent interference effects between the beams.
Absorption imaging is performed along the $-y$ direction, using the $F=2 \rightarrow 3'$ cycling transition after a brief \SI{0.5}{ms} time-of-flight expansion.

\subsection{Classical Trajectories}
\label{suppsec:2}
Here we present the analysis of the classical trajectories scattered by the Gaussian potential \cref{eq:U}.
Sufficiently far from the potential, $U(x, z) \approx 0$ and the trajectories will be parabolic due to the constant downward acceleration $a_z$:
\begin{gather*}
  \vect{q}(t) = \begin{pmatrix}
    x(t)\\
    z(t)
  \end{pmatrix}
  =
  \begin{pmatrix}
    x_0(x_i) + v_0 t \sin\theta(x_i) \phantom{{} - \frac{a_zt^2}{2} - h}\\
    z_0(x_i) - v_0 t \cos\theta(x_i) - \frac{a_zt^2}{2} - h
  \end{pmatrix}.
\end{gather*}
Here $v_0 = \sqrt{2a_z h}$ is the speed of particles falling from the injection site without the potential.
In the limit of a zero-range potential, $\sigma \rightarrow 0$, one has $x_0(x_i) \rightarrow 0$, $z_0(x_i) \rightarrow 0$.
In this limit, $\theta(x_i)$ is the scattering angle at time $t=0$ when the particle hits the potential, and is the only parameter that depends on the impact parameter $x_i$.
Deviations from these values characterize the finite-size effects of the potential, and must be calculated numerically.

The effects of imaging after a time-of-flight expansion of $t_I = \SI{0.5}{ms}$ can be included by extending the trajectories from time $t$ to time $t_f = t+t_I$ under the reduced acceleration $g$:
\begin{gather*}
  \vect{q}(t_f,t_I)
  =
  \begin{pmatrix}
    x_0 + v_0 t_f \sin\theta \phantom{{}- \frac{a_zt_f^2}{2} + \frac{(a_z - g)t_I^2}{2} - h}\\
    z_0 - v_0 t_f \cos\theta - \frac{a_z t_f^2}{2} -h + \frac{(a_z - g)t_I^2}{2}
  \end{pmatrix}.
\end{gather*}
The effect of imaging is, thus, to shift the trajectories after scattering vertically.
The classical trajectories can be found by eliminating $t_f$, and are inverted parabola
\begin{gather}
  z(x) = z_* - \frac{(x-x_*)^2}{4 h\sin^2\theta}
\end{gather}
with maxima at $(x_*, z_*)$:
\begin{align*}
  x_* &= x_0 - h\sin 2\theta, &
  z_* &= z_0 - h \sin^2\theta + \frac{(a_z - g)t_I^2}{2}.
\end{align*}
Note that if the scattering is such that the maximum scattering angle
\begin{gather}
  \theta_c = \max_{x_i} \theta
\end{gather}
is less than \SI{90}{\degree} ($\abs{\theta_c} < \pi/2$), then the trajectory where $\theta = \theta_c$ will have the widest parabola.
This trajectory will then eventually overtake all other trajectories and corresponds to the limit of trajectories as $\theta \rightarrow \theta_c$, and will lie along a caustic, corresponding to a singularity in $\det{\mat{J}} = 0$ [\cref{eq:detJ}] with vanishing partials $\partial/\partial x_i$.

To perform the analysis in \cref{fig:theta} of the case where $0 < \epsilon < 1$, we locate the outer caustic in each image and then fit the parabolic scattering trajectory to these, extracting the angle $\theta_c(\varepsilon)$.
This analysis was repeated for three experimental runs of the same parameters, where the resulting average and estimated error are reported in \cref{fig:theta}.

All data has been corrected for a small camera tilt of $\theta_{\mathrm{cam}}=\SI{0.81}{\degree}$, which was measured using a falling \gls{BEC} from a tightly confined trap.

Neglecting the effects of acceleration \emph{during} the scattering, $\theta_c(\varepsilon)$ depends only on the dimensionless ratio $\varepsilon$.
In particular, $\theta_c$ is independent of the waist $\sigma$ of the potential, which only changes which impact parameter $x_i(\varepsilon, \sigma)$ scatters maximally.
Interestingly, as shown in \cref{fig:theta}, scattering from a Gaussian potential is somewhat peculiar in that the maximum scattering angle for $\num{-2.59} \lessapprox \varepsilon \lessapprox 1$ is less than \SI{90}{\degree}.
This is the case for all examples considered here except for \cref{fig:GaussShock-r4} which has $\epsilon > 1$ and hence $\theta_c = 0$ for the bouncing trajectories.
In this limit, the transition from $\varepsilon < 1$ to $\varepsilon > 1$ is discontinuous with the sudden disappearance of the attached oblique-shock--like caustic and the appearance of the detached bow-shock--like caustic.

This discontinuity is slightly softened by the fact that, during the scattering, the acceleration $a_z$ is still present.
In this case, $\theta_c(\varepsilon, \varsigma)$ depends on the two dimensionless ratios
\begin{gather}
  \varepsilon = \frac{U_0}{ma_zh}, \qquad \varsigma = \frac{\sigma}{h}.
\end{gather}
The effective potential for the central trajectory $x_i=0$ is thus:
\begin{gather}
  \frac{V(z)}{mah} = \tilde{V}(\tilde{z}) = \tilde{z} + \varepsilon e^{-2\tilde{z}^2/\varsigma^2},
  \qquad
  \tilde{z} = \frac{z+h}{h}.
\end{gather}
This particle will bounce if $\tilde{V}_{\max}(\tilde{z}) \geq \tilde{V}(1) \approx 1$, where the latter approximation for all values of $\varsigma$ we consider for the repulsive potentials.
The explicit solution to this optimization problem is:
\begin{gather}
  \tilde{z}_c = \frac{1 - \sqrt{1-\varsigma^2}}{2}, \qquad
  \varepsilon = \frac{\varsigma^2}{4\tilde{z}_c}e^{2\tilde{z}_c^2/\varsigma^2}.
\end{gather}
With our parameters, $\varsigma \exclude{= \SI{11.3}{µm}/\SI{78}{µm}} = 0.14$, corresponding to a transition region of $\num{0.997} \leq \varepsilon < 1$.
Frame \cref{fig:GaussShock-r3}, for example, shows faint signatures of both features, and thus sits right at this transition.

We can also consider the slices of constant $t_f$ corresponding to the dark bands in \cref{fig:FFtrace}:
\begin{gather*}
  x(x_i) = x_0(x_i) + v_0t_f\sin\theta(x_i)\\
  z(x_i) = z_0(x_i) - v_0t_f\cos\theta(x_i) - \tfrac{a_zt_f^2}{2} - h + \tfrac{(a_z-g)t_I^2}{2}.
\end{gather*}
In the zero-range limit $x_0, z_0 \rightarrow 0$ no longer depend on $x_i$ so we can solve for $\theta(x_i)$ to obtain:
\begin{gather}
  z(x) \approx - \sqrt{v_0^2t_f^2-x^2} - \tfrac{a_zt_f^2}{2} - h + \tfrac{(a_z-g)t_I^2}{2}.
\end{gather}
These are circular arcs of radius $v_0t_f$ centered at height $-h + (a_z-g)t_I^2/2 - a_zt_f^2/2$, extending through $\theta \in [-\theta_c, \theta_c]$.

Our numerical results do not make these approximations.
The classical trajectories are found by integrating the classical equations of motion with the physical potential, including the expansion time.
The agreement between these, however, allows us to assert that the quantitative corrections from the acceleration during scattering are small (sub percent).

\section{Code Availability}
All relevant code used for numerical studies in this work is available from the corresponding authors upon reasonable request.
Additional code for visualizing the 3\textsc{d} caustics is available from~\cite{gitlab:catastrophe_atom_optics}.

\section{Data Availability}
The data used to generate the figures in the manuscript and accompanying code are available under \href{https://osf.io/kdm9s/}{osf.io/kdm9s/}~\cite{osf:catastrophe_atom_optics}.
Additional experimental and numerical data sets in this work will be made available from the corresponding authors upon reasonable request.

\newif\ifusebbl
\usebbltrue   
\ifusebbl

\else
  \bibliography{macros,ThesisRef,local}

\begin{thebibliography}{67}%
\makeatletter
\providecommand \@ifxundefined [1]{%
 \@ifx{#1\undefined}
}%
\providecommand \@ifnum [1]{%
 \ifnum #1\expandafter \@firstoftwo
 \else \expandafter \@secondoftwo
 \fi
}%
\providecommand \@ifx [1]{%
 \ifx #1\expandafter \@firstoftwo
 \else \expandafter \@secondoftwo
 \fi
}%
\providecommand \natexlab [1]{#1}%
\providecommand \enquote  [1]{``#1''}%
\providecommand \bibnamefont  [1]{#1}%
\providecommand \bibfnamefont [1]{#1}%
\providecommand \citenamefont [1]{#1}%
\providecommand \href@noop [0]{\@secondoftwo}%
\providecommand \href [0]{\begingroup \@sanitize@url \@href}%
\providecommand \@href[1]{\@@startlink{#1}\@@href}%
\providecommand \@@href[1]{\endgroup#1\@@endlink}%
\providecommand \@sanitize@url [0]{\catcode `\\12\catcode `\$12\catcode
  `\&12\catcode `\#12\catcode `\^12\catcode `\_12\catcode `\%12\relax}%
\providecommand \@@startlink[1]{}%
\providecommand \@@endlink[0]{}%
\providecommand \url  [0]{\begingroup\@sanitize@url \@url }%
\providecommand \@url [1]{\endgroup\@href {#1}{\urlprefix }}%
\providecommand \urlprefix  [0]{URL }%
\providecommand \Eprint [0]{\href }%
\providecommand \doibase [0]{https://doi.org/}%
\providecommand \selectlanguage [0]{\@gobble}%
\providecommand \bibinfo  [0]{\@secondoftwo}%
\providecommand \bibfield  [0]{\@secondoftwo}%
\providecommand \translation [1]{[#1]}%
\providecommand \BibitemOpen [0]{}%
\providecommand \bibitemStop [0]{}%
\providecommand \bibitemNoStop [0]{.\EOS\space}%
\providecommand \EOS [0]{\spacefactor3000\relax}%
\providecommand \BibitemShut  [1]{\csname bibitem#1\endcsname}%
\let\auto@bib@innerbib\@empty
\bibitem [{\citenamefont {Berry}\ and\ \citenamefont
  {Upstill}(1980)}]{Berry1980}%
  \BibitemOpen
  \bibfield  {author} {\bibinfo {author} {\bibfnamefont {M.~V.}\ \bibnamefont
  {Berry}}\ and\ \bibinfo {author} {\bibfnamefont {C.}~\bibnamefont
  {Upstill}},\ }\bibfield  {title} {\bibinfo {title} {Catastrophe optics:
  Morphologies of caustics and their diffraction patterns},\ }\href
  {https://doi.org/10.1016/S0079-6638(08)70215-4} {\bibfield  {journal}
  {\bibinfo  {journal} {Progress in Optics}\ }\textbf {\bibinfo {volume}
  {18}},\ \bibinfo {pages} {257} (\bibinfo {year} {1980})}\BibitemShut
  {NoStop}%
\bibitem [{\citenamefont {Nye}(1999)}]{Nye1999}%
  \BibitemOpen
  \bibfield  {author} {\bibinfo {author} {\bibfnamefont {J.~F.}\ \bibnamefont
  {Nye}},\ }\href {https://doi.org/10.1119/1.19543} {\emph {\bibinfo {title}
  {Natural Focusing and Fine Structure of Light: Caustics and Wave
  Dislocations}}}\ (\bibinfo  {publisher} {Institute of Physics Publishing
  Ltd},\ \bibinfo {year} {1999})\BibitemShut {NoStop}%
\bibitem [{\citenamefont {Born}\ and\ \citenamefont {Wolf}(1999)}]{Born1999}%
  \BibitemOpen
  \bibfield  {author} {\bibinfo {author} {\bibfnamefont {M.}~\bibnamefont
  {Born}}\ and\ \bibinfo {author} {\bibfnamefont {E.}~\bibnamefont {Wolf}},\
  }\href {https://doi.org/10.1017/CBO9781139644181} {\emph {\bibinfo {title}
  {{Priciples of Optics: Electromagnetic Theory of Propagation, Interference
  and Diffraction of Light}}}},\ \bibinfo {edition} {7th}\ ed.\ (\bibinfo
  {publisher} {Cambridge University Press},\ \bibinfo {year}
  {1999})\BibitemShut {NoStop}%
\bibitem [{\citenamefont {Weinstein}(1969)}]{Weinstein1969}%
  \BibitemOpen
  \bibfield  {author} {\bibinfo {author} {\bibfnamefont {L.~A.}\ \bibnamefont
  {Weinstein}},\ }\href {https://doi.org/10.1119/1.1976211} {\emph {\bibinfo
  {title} {Open resonators and open waveguides}}},\ edited by\ \bibinfo
  {editor} {\bibfnamefont {P.}~\bibnamefont {Beckmann}},\ Golem series in
  electromagnetics\ (\bibinfo  {publisher} {Golem Press},\ \bibinfo {year}
  {1969})\BibitemShut {NoStop}%
\bibitem [{\citenamefont {Berry}(1981)}]{Berry1981}%
  \BibitemOpen
  \bibfield  {author} {\bibinfo {author} {\bibfnamefont {M.~V.}\ \bibnamefont
  {Berry}},\ }\bibfield  {title} {\bibinfo {title} {Singularities in waves and
  rays},\ }in\ \href@noop {} {\emph {\bibinfo {booktitle} {Physics of
  Defects}}},\ \bibinfo {editor} {edited by\ \bibinfo {editor} {\bibfnamefont
  {R.}~\bibnamefont {Balian}}, \bibinfo {editor} {\bibfnamefont
  {M.}~\bibnamefont {Kl{\'e}man}},\ and\ \bibinfo {editor} {\bibfnamefont
  {J.-P.}\ \bibnamefont {Poirier}}}\ (\bibinfo  {publisher} {North-Holland,
  Amsterdam},\ \bibinfo {year} {1981})\ pp.\ \bibinfo {pages}
  {453--543}\BibitemShut {NoStop}%
\bibitem [{\citenamefont {Marston}\ and\ \citenamefont
  {Trinh}(1984)}]{Marston1984}%
  \BibitemOpen
  \bibfield  {author} {\bibinfo {author} {\bibfnamefont {P.~L.}\ \bibnamefont
  {Marston}}\ and\ \bibinfo {author} {\bibfnamefont {E.~H.}\ \bibnamefont
  {Trinh}},\ }\bibfield  {title} {\bibinfo {title} {Hyperbolic umbilic
  diffraction catastrophe and rainbow scattering from spheroidal drops},\
  }\href {https://doi.org/10.1038/312529a0} {\bibfield  {journal} {\bibinfo
  {journal} {Nature}\ }\textbf {\bibinfo {volume} {312}},\ \bibinfo {pages}
  {529} (\bibinfo {year} {1984})}\BibitemShut {NoStop}%
\bibitem [{\citenamefont {Kaduchak}\ and\ \citenamefont
  {Marston}(1994)}]{Kaduchak1994}%
  \BibitemOpen
  \bibfield  {author} {\bibinfo {author} {\bibfnamefont {G.}~\bibnamefont
  {Kaduchak}}\ and\ \bibinfo {author} {\bibfnamefont {P.~L.}\ \bibnamefont
  {Marston}},\ }\bibfield  {title} {\bibinfo {title} {Hyperbolic umbilic and
  $e_{6}$ diffraction catastrophes associated with the secondary rainbow of
  oblate water drops: observations with laser illumination},\ }\href
  {https://doi.org/10.1364/AO.33.004697} {\bibfield  {journal} {\bibinfo
  {journal} {Applied Optics}\ }\textbf {\bibinfo {volume} {33}},\ \bibinfo
  {pages} {4697} (\bibinfo {year} {1994})}\BibitemShut {NoStop}%
\bibitem [{\citenamefont {Borghi}(2016)}]{Borghi2016}%
  \BibitemOpen
  \bibfield  {author} {\bibinfo {author} {\bibfnamefont {R.}~\bibnamefont
  {Borghi}},\ }\bibfield  {title} {\bibinfo {title} {Catastrophe optics of
  sharp edge diffraction},\ }\href {https://doi.org/10.1364/OL.41.003114}
  {\bibfield  {journal} {\bibinfo  {journal} {Opt. Lett.}\ }\textbf {\bibinfo
  {volume} {41}},\ \bibinfo {pages} {3114} (\bibinfo {year}
  {2016})}\BibitemShut {NoStop}%
\bibitem [{\citenamefont {Thom}(1975)}]{Thom:1975}%
  \BibitemOpen
  \bibfield  {author} {\bibinfo {author} {\bibfnamefont {R.}~\bibnamefont
  {Thom}},\ }\href {https://doi.org/10.2307/2065331} {\emph {\bibinfo {title}
  {Structural stability and morphogenesis; an outline of a general theory of
  models}}},\ \bibinfo {edition} {1st}\ ed.\ (\bibinfo  {publisher} {W. A.
  Benjamin},\ \bibinfo {address} {Reading, Massachusetts},\ \bibinfo {year}
  {1975})\ p.\ \bibinfo {pages} {348},\ \bibinfo {note} {translated from the
  French ed., as updated by the author, by D. H. Fowler. With a foreword by C.
  H. Waddington}\BibitemShut {NoStop}%
\bibitem [{\citenamefont {Arnol'd}(1992)}]{Arnold:1986}%
  \BibitemOpen
  \bibfield  {author} {\bibinfo {author} {\bibfnamefont {V.~I.}\ \bibnamefont
  {Arnol'd}},\ }\href {https://doi.org/10.1007/978-3-642-58124-3} {\emph
  {\bibinfo {title} {Catastrophe Theory}}}\ (\bibinfo  {publisher}
  {Springer-Verlag},\ \bibinfo {year} {1992})\BibitemShut {NoStop}%
\bibitem [{\citenamefont {O'Dell}(2012)}]{ODell:2012}%
  \BibitemOpen
  \bibfield  {author} {\bibinfo {author} {\bibfnamefont {D.~H.~J.}\
  \bibnamefont {O'Dell}},\ }\bibfield  {title} {\bibinfo {title} {Quantum
  catastrophes and ergodicity in the dynamics of bosonic josephson junctions},\
  }\href {https://doi.org/10.1103/PhysRevLett.109.150406} {\bibfield  {journal}
  {\bibinfo  {journal} {Phys. Rev. Lett.}\ }\textbf {\bibinfo {volume} {109}},\
  \bibinfo {pages} {150406} (\bibinfo {year} {2012})}\BibitemShut {NoStop}%
\bibitem [{\citenamefont {Mumford}\ \emph {et~al.}(2017)\citenamefont
  {Mumford}, \citenamefont {Kirkby},\ and\ \citenamefont
  {O'Dell}}]{Mumford:2017}%
  \BibitemOpen
  \bibfield  {author} {\bibinfo {author} {\bibfnamefont {J.}~\bibnamefont
  {Mumford}}, \bibinfo {author} {\bibfnamefont {W.}~\bibnamefont {Kirkby}},\
  and\ \bibinfo {author} {\bibfnamefont {D.~H.~J.}\ \bibnamefont {O'Dell}},\
  }\bibfield  {title} {\bibinfo {title} {Catastrophes in non-equilibrium
  many-particle wave functions: universality and critical scaling},\ }\href
  {https://doi.org/10.1088/1361-6455/aa56af} {\bibfield  {journal} {\bibinfo
  {journal} {J. Phys. B: At. Mol. Opt. Phys.}\ }\textbf {\bibinfo {volume}
  {50}},\ \bibinfo {pages} {044005} (\bibinfo {year} {2017})}\BibitemShut
  {NoStop}%
\bibitem [{\citenamefont {Mumford}\ \emph {et~al.}(2019)\citenamefont
  {Mumford}, \citenamefont {Turner}, \citenamefont {Sprung},\ and\
  \citenamefont {O'Dell}}]{Mumford:2019}%
  \BibitemOpen
  \bibfield  {author} {\bibinfo {author} {\bibfnamefont {J.}~\bibnamefont
  {Mumford}}, \bibinfo {author} {\bibfnamefont {E.}~\bibnamefont {Turner}},
  \bibinfo {author} {\bibfnamefont {D.~W.~L.}\ \bibnamefont {Sprung}},\ and\
  \bibinfo {author} {\bibfnamefont {D.~H.~J.}\ \bibnamefont {O'Dell}},\
  }\bibfield  {title} {\bibinfo {title} {Quantum spin dynamics in fock space
  following quenches: Caustics and vortices},\ }\href
  {https://doi.org/10.1103/PhysRevLett.122.170402} {\bibfield  {journal}
  {\bibinfo  {journal} {Phys. Rev. Lett.}\ }\textbf {\bibinfo {volume} {122}},\
  \bibinfo {pages} {170402} (\bibinfo {year} {2019})}\BibitemShut {NoStop}%
\bibitem [{\citenamefont {{Da Silveira}}(1973)}]{DaSilveira:1973}%
  \BibitemOpen
  \bibfield  {author} {\bibinfo {author} {\bibfnamefont {R.}~\bibnamefont {{Da
  Silveira}}},\ }\bibfield  {title} {\bibinfo {title} {Rainbow interference
  effects in heavy ion elastic scattering},\ }\href
  {https://doi.org/https://doi.org/10.1016/0370-2693(73)90184-6} {\bibfield
  {journal} {\bibinfo  {journal} {Physics Letters B}\ }\textbf {\bibinfo
  {volume} {45}},\ \bibinfo {pages} {211} (\bibinfo {year} {1973})}\BibitemShut
  {NoStop}%
\bibitem [{\citenamefont {Hasse}\ \emph {et~al.}(1996)\citenamefont {Hasse},
  \citenamefont {Kriele},\ and\ \citenamefont {Perlick}}]{Hasse_1996}%
  \BibitemOpen
  \bibfield  {author} {\bibinfo {author} {\bibfnamefont {W.}~\bibnamefont
  {Hasse}}, \bibinfo {author} {\bibfnamefont {M.}~\bibnamefont {Kriele}},\ and\
  \bibinfo {author} {\bibfnamefont {V.}~\bibnamefont {Perlick}},\ }\bibfield
  {title} {\bibinfo {title} {Caustics of wavefronts in general relativity},\
  }\href {https://doi.org/10.1088/0264-9381/13/5/027} {\bibfield  {journal}
  {\bibinfo  {journal} {Classical and Quantum Gravity}\ }\textbf {\bibinfo
  {volume} {13}},\ \bibinfo {pages} {1161} (\bibinfo {year}
  {1996})}\BibitemShut {NoStop}%
\bibitem [{\citenamefont {Oliva}\ \emph {et~al.}(1981)\citenamefont {Oliva},
  \citenamefont {Peters},\ and\ \citenamefont {Murthy}}]{Oliva:1981}%
  \BibitemOpen
  \bibfield  {author} {\bibinfo {author} {\bibfnamefont {T.~A.}\ \bibnamefont
  {Oliva}}, \bibinfo {author} {\bibfnamefont {M.~H.}\ \bibnamefont {Peters}},\
  and\ \bibinfo {author} {\bibfnamefont {H.~S.~K.}\ \bibnamefont {Murthy}},\
  }\bibfield  {title} {\bibinfo {title} {A preliminary empirical test of a cusp
  catastrophe model in the social sciences},\ }\href
  {https://doi.org/10.1002/bs.3830260208} {\bibfield  {journal} {\bibinfo
  {journal} {Behavioral Science}\ }\textbf {\bibinfo {volume} {26}},\ \bibinfo
  {pages} {153} (\bibinfo {year} {1981})}\BibitemShut {NoStop}%
\bibitem [{\citenamefont {Carricato}\ \emph {et~al.}(2002)\citenamefont
  {Carricato}, \citenamefont {Duffy},\ and\ \citenamefont
  {Parenti-Castelli}}]{Carricato:2002}%
  \BibitemOpen
  \bibfield  {author} {\bibinfo {author} {\bibfnamefont {M.}~\bibnamefont
  {Carricato}}, \bibinfo {author} {\bibfnamefont {J.}~\bibnamefont {Duffy}},\
  and\ \bibinfo {author} {\bibfnamefont {V.}~\bibnamefont {Parenti-Castelli}},\
  }\bibfield  {title} {\bibinfo {title} {Catastrophe analysis of a planar
  system with flexural pivots},\ }\href
  {https://doi.org/10.1016/s0094-114x(02)00012-5} {\bibfield  {journal}
  {\bibinfo  {journal} {Mechanism and Machine Theory}\ }\textbf {\bibinfo
  {volume} {37}},\ \bibinfo {pages} {693} (\bibinfo {year} {2002})}\BibitemShut
  {NoStop}%
\bibitem [{\citenamefont {Petersen}\ \emph {et~al.}(2013)\citenamefont
  {Petersen}, \citenamefont {Weyland}, \citenamefont {Paganin}, \citenamefont
  {Simula}, \citenamefont {Eastwood},\ and\ \citenamefont
  {Morgan}}]{Petersen:2013}%
  \BibitemOpen
  \bibfield  {author} {\bibinfo {author} {\bibfnamefont {T.~C.}\ \bibnamefont
  {Petersen}}, \bibinfo {author} {\bibfnamefont {M.}~\bibnamefont {Weyland}},
  \bibinfo {author} {\bibfnamefont {D.~M.}\ \bibnamefont {Paganin}}, \bibinfo
  {author} {\bibfnamefont {T.~P.}\ \bibnamefont {Simula}}, \bibinfo {author}
  {\bibfnamefont {S.~A.}\ \bibnamefont {Eastwood}},\ and\ \bibinfo {author}
  {\bibfnamefont {M.~J.}\ \bibnamefont {Morgan}},\ }\bibfield  {title}
  {\bibinfo {title} {Electron vortex production and control using aberration
  induced diffraction catastrophes},\ }\href
  {https://doi.org/10.1103/PhysRevLett.110.033901} {\bibfield  {journal}
  {\bibinfo  {journal} {Phys. Rev. Lett.}\ }\textbf {\bibinfo {volume} {110}},\
  \bibinfo {pages} {033901} (\bibinfo {year} {2013})}\BibitemShut {NoStop}%
\bibitem [{\citenamefont {Shimizu}\ and\ \citenamefont {inci
  Fujita}(2002)}]{Shimizu2002}%
  \BibitemOpen
  \bibfield  {author} {\bibinfo {author} {\bibfnamefont {F.}~\bibnamefont
  {Shimizu}}\ and\ \bibinfo {author} {\bibfnamefont {J.}~\bibnamefont {inci
  Fujita}},\ }\bibfield  {title} {\bibinfo {title} {Reflection-type hologram
  for atoms},\ }\href {https://doi.org/10.1103/PhysRevLett.88.123201}
  {\bibfield  {journal} {\bibinfo  {journal} {Phys. Rev. Lett.}\ }\textbf
  {\bibinfo {volume} {88}},\ \bibinfo {pages} {123201} (\bibinfo {year}
  {2002})}\BibitemShut {NoStop}%
\bibitem [{\citenamefont {Balykin}\ \emph {et~al.}(2003)\citenamefont
  {Balykin}, \citenamefont {Klimov},\ and\ \citenamefont
  {Letokhov}}]{Balykin2003}%
  \BibitemOpen
  \bibfield  {author} {\bibinfo {author} {\bibfnamefont {V.~I.}\ \bibnamefont
  {Balykin}}, \bibinfo {author} {\bibfnamefont {V.~V.}\ \bibnamefont
  {Klimov}},\ and\ \bibinfo {author} {\bibfnamefont {V.~S.}\ \bibnamefont
  {Letokhov}},\ }\bibfield  {title} {\bibinfo {title} {Atom nanooptics based on
  photon dots and photon holes},\ }\href {https://doi.org/10.1134/1.1609567}
  {\bibfield  {journal} {\bibinfo  {journal} {JETP Letters}\ }\textbf {\bibinfo
  {volume} {78}},\ \bibinfo {pages} {8} (\bibinfo {year} {2003})}\BibitemShut
  {NoStop}%
\bibitem [{\citenamefont {Oberst}\ \emph {et~al.}(2005)\citenamefont {Oberst},
  \citenamefont {Kouznetsov}, \citenamefont {Shimizu}, \citenamefont {inci
  Fujita},\ and\ \citenamefont {Shimizu}}]{Oberst2005}%
  \BibitemOpen
  \bibfield  {author} {\bibinfo {author} {\bibfnamefont {H.}~\bibnamefont
  {Oberst}}, \bibinfo {author} {\bibfnamefont {D.}~\bibnamefont {Kouznetsov}},
  \bibinfo {author} {\bibfnamefont {K.}~\bibnamefont {Shimizu}}, \bibinfo
  {author} {\bibfnamefont {J.}~\bibnamefont {inci Fujita}},\ and\ \bibinfo
  {author} {\bibfnamefont {F.}~\bibnamefont {Shimizu}},\ }\bibfield  {title}
  {\bibinfo {title} {Fresnel diffraction mirror for an atomic wave},\ }\href
  {https://doi.org/10.1103/PhysRevLett.94.013203} {\bibfield  {journal}
  {\bibinfo  {journal} {Phys. Rev. Lett.}\ }\textbf {\bibinfo {volume} {94}},\
  \bibinfo {pages} {013203} (\bibinfo {year} {2005})}\BibitemShut {NoStop}%
\bibitem [{\citenamefont {Kouznetsov}\ \emph {et~al.}(2006)\citenamefont
  {Kouznetsov}, \citenamefont {Oberst}, \citenamefont {Neumann}, \citenamefont
  {Kuznetsova}, \citenamefont {Shimizu}, \citenamefont {Bisson}, \citenamefont
  {Ueda},\ and\ \citenamefont {Brueck}}]{Kouznetsov2006}%
  \BibitemOpen
  \bibfield  {author} {\bibinfo {author} {\bibfnamefont {D.}~\bibnamefont
  {Kouznetsov}}, \bibinfo {author} {\bibfnamefont {H.}~\bibnamefont {Oberst}},
  \bibinfo {author} {\bibfnamefont {A.}~\bibnamefont {Neumann}}, \bibinfo
  {author} {\bibfnamefont {Y.}~\bibnamefont {Kuznetsova}}, \bibinfo {author}
  {\bibfnamefont {K.}~\bibnamefont {Shimizu}}, \bibinfo {author} {\bibfnamefont
  {J.-F.}\ \bibnamefont {Bisson}}, \bibinfo {author} {\bibfnamefont
  {K.}~\bibnamefont {Ueda}},\ and\ \bibinfo {author} {\bibfnamefont {S.~R.~J.}\
  \bibnamefont {Brueck}},\ }\bibfield  {title} {\bibinfo {title} {Ridged atomic
  mirrors and atomic nanoscope},\ }\href
  {https://doi.org/10.1088/0953-4075/39/7/005} {\bibfield  {journal} {\bibinfo
  {journal} {J. Phys. B: At. Mol. Opt. Phys.}\ }\textbf {\bibinfo {volume}
  {39}},\ \bibinfo {pages} {1605} (\bibinfo {year} {2006})}\BibitemShut
  {NoStop}%
\bibitem [{\citenamefont {Rohwedder}(2007)}]{Rohwedder2007}%
  \BibitemOpen
  \bibfield  {author} {\bibinfo {author} {\bibfnamefont {B.}~\bibnamefont
  {Rohwedder}},\ }\bibfield  {title} {\bibinfo {title} {Resource letter aon-1:
  Atom optics, a tool for nanofabrication},\ }\href
  {https://doi.org/10.1119/1.2673209} {\bibfield  {journal} {\bibinfo
  {journal} {American Journal of Physics}\ }\textbf {\bibinfo {volume} {75}},\
  \bibinfo {pages} {394} (\bibinfo {year} {2007})}\BibitemShut {NoStop}%
\bibitem [{\citenamefont {Cronin}\ \emph {et~al.}(2009)\citenamefont {Cronin},
  \citenamefont {Schmiedmayer},\ and\ \citenamefont {Pritchard}}]{Cronin2009}%
  \BibitemOpen
  \bibfield  {author} {\bibinfo {author} {\bibfnamefont {A.~D.}\ \bibnamefont
  {Cronin}}, \bibinfo {author} {\bibfnamefont {J.}~\bibnamefont
  {Schmiedmayer}},\ and\ \bibinfo {author} {\bibfnamefont {D.~E.}\ \bibnamefont
  {Pritchard}},\ }\bibfield  {title} {\bibinfo {title} {Optics and
  interferometry with atoms and molecules},\ }\href
  {https://doi.org/10.1103/RevModPhys.81.1051} {\bibfield  {journal} {\bibinfo
  {journal} {Rev. Mod. Phys}\ }\textbf {\bibinfo {volume} {81}},\ \bibinfo
  {pages} {1051} (\bibinfo {year} {2009})}\BibitemShut {NoStop}%
\bibitem [{\citenamefont {Dall}\ \emph {et~al.}(2010)\citenamefont {Dall},
  \citenamefont {Hodgman}, \citenamefont {Johnsson}, \citenamefont {Baldwin},\
  and\ \citenamefont {Truscott}}]{Dall:2010}%
  \BibitemOpen
  \bibfield  {author} {\bibinfo {author} {\bibfnamefont {R.~G.}\ \bibnamefont
  {Dall}}, \bibinfo {author} {\bibfnamefont {S.~S.}\ \bibnamefont {Hodgman}},
  \bibinfo {author} {\bibfnamefont {M.~T.}\ \bibnamefont {Johnsson}}, \bibinfo
  {author} {\bibfnamefont {K.~G.~H.}\ \bibnamefont {Baldwin}},\ and\ \bibinfo
  {author} {\bibfnamefont {A.~G.}\ \bibnamefont {Truscott}},\ }\bibfield
  {title} {\bibinfo {title} {Transverse mode imaging of guided matter waves},\
  }\href {https://doi.org/10.1103/PhysRevA.81.011602} {\bibfield  {journal}
  {\bibinfo  {journal} {Physical Review A}\ }\textbf {\bibinfo {volume} {81}},\
  \bibinfo {pages} {011602} (\bibinfo {year} {2010})}\BibitemShut {NoStop}%
\bibitem [{\citenamefont {Simula}\ \emph {et~al.}(2013)\citenamefont {Simula},
  \citenamefont {Petersen},\ and\ \citenamefont {Paganin}}]{Simula2013}%
  \BibitemOpen
  \bibfield  {author} {\bibinfo {author} {\bibfnamefont {T.~P.}\ \bibnamefont
  {Simula}}, \bibinfo {author} {\bibfnamefont {T.~C.}\ \bibnamefont
  {Petersen}},\ and\ \bibinfo {author} {\bibfnamefont {D.~M.}\ \bibnamefont
  {Paganin}},\ }\bibfield  {title} {\bibinfo {title} {Diffraction catastrophes
  threaded by quantized vortex skeletons caused by atom-optical aberrations
  induced in trapped {B}ose-{E}instein condensates},\ }\href
  {https://doi.org/10.1103/PhysRevA.88.043626} {\bibfield  {journal} {\bibinfo
  {journal} {Phys. Rev. A}\ }\textbf {\bibinfo {volume} {88}},\ \bibinfo
  {pages} {043626} (\bibinfo {year} {2013})}\BibitemShut {NoStop}%
\bibitem [{\citenamefont {Rooijakkers}\ \emph {et~al.}(2003)\citenamefont
  {Rooijakkers}, \citenamefont {Wu}, \citenamefont {Striehl}, \citenamefont
  {Vengalattore},\ and\ \citenamefont {Prentiss}}]{Rooijakkers:2003}%
  \BibitemOpen
  \bibfield  {author} {\bibinfo {author} {\bibfnamefont {W.}~\bibnamefont
  {Rooijakkers}}, \bibinfo {author} {\bibfnamefont {S.}~\bibnamefont {Wu}},
  \bibinfo {author} {\bibfnamefont {P.}~\bibnamefont {Striehl}}, \bibinfo
  {author} {\bibfnamefont {M.}~\bibnamefont {Vengalattore}},\ and\ \bibinfo
  {author} {\bibfnamefont {M.}~\bibnamefont {Prentiss}},\ }\bibfield  {title}
  {\bibinfo {title} {Observation of caustics in the trajectories of cold atoms
  in a linear magnetic potential},\ }\href
  {https://doi.org/10.1103/PhysRevA.68.063412} {\bibfield  {journal} {\bibinfo
  {journal} {Phys. Rev. A}\ }\textbf {\bibinfo {volume} {68}},\ \bibinfo
  {pages} {063412} (\bibinfo {year} {2003})}\BibitemShut {NoStop}%
\bibitem [{\citenamefont {Rosenblum}\ \emph {et~al.}(2014)\citenamefont
  {Rosenblum}, \citenamefont {Bechler}, \citenamefont {Shomroni}, \citenamefont
  {Kaner}, \citenamefont {Arusi-Parpar}, \citenamefont {Raz},\ and\
  \citenamefont {Dayan}}]{Rosenblum2014}%
  \BibitemOpen
  \bibfield  {author} {\bibinfo {author} {\bibfnamefont {S.}~\bibnamefont
  {Rosenblum}}, \bibinfo {author} {\bibfnamefont {O.}~\bibnamefont {Bechler}},
  \bibinfo {author} {\bibfnamefont {I.}~\bibnamefont {Shomroni}}, \bibinfo
  {author} {\bibfnamefont {R.}~\bibnamefont {Kaner}}, \bibinfo {author}
  {\bibfnamefont {T.}~\bibnamefont {Arusi-Parpar}}, \bibinfo {author}
  {\bibfnamefont {O.}~\bibnamefont {Raz}},\ and\ \bibinfo {author}
  {\bibfnamefont {B.}~\bibnamefont {Dayan}},\ }\bibfield  {title} {\bibinfo
  {title} {Demonstration of fold and cusp catastrophes in an atomic cloud
  reflected from an optical barrier in the presence of gravity},\ }\href
  {https://doi.org/10.1103/PhysRevLett.112.120403} {\bibfield  {journal}
  {\bibinfo  {journal} {Phys. Rev. Lett.}\ }\textbf {\bibinfo {volume} {112}},\
  \bibinfo {pages} {120403} (\bibinfo {year} {2014})}\BibitemShut {NoStop}%
\bibitem [{\citenamefont {O{\textquotesingle}Dell}(2001)}]{ODell:2001}%
  \BibitemOpen
  \bibfield  {author} {\bibinfo {author} {\bibfnamefont {D.~H.~J.}\
  \bibnamefont {O{\textquotesingle}Dell}},\ }\bibfield  {title} {\bibinfo
  {title} {Dynamical diffraction in sinusoidal potentials: uniform
  approximations for mathieu functions},\ }\href
  {https://doi.org/10.1088/0305-4470/34/18/316} {\bibfield  {journal} {\bibinfo
   {journal} {Journal of Physics A: Mathematical and General}\ }\textbf
  {\bibinfo {volume} {34}},\ \bibinfo {pages} {3897} (\bibinfo {year}
  {2001})}\BibitemShut {NoStop}%
\bibitem [{\citenamefont {Huckans}\ \emph {et~al.}(2009)\citenamefont
  {Huckans}, \citenamefont {Spielman}, \citenamefont {Tolra}, \citenamefont
  {Phillips},\ and\ \citenamefont {Porto}}]{Huckans:2009}%
  \BibitemOpen
  \bibfield  {author} {\bibinfo {author} {\bibfnamefont {J.~H.}\ \bibnamefont
  {Huckans}}, \bibinfo {author} {\bibfnamefont {I.~B.}\ \bibnamefont
  {Spielman}}, \bibinfo {author} {\bibfnamefont {B.~L.}\ \bibnamefont {Tolra}},
  \bibinfo {author} {\bibfnamefont {W.~D.}\ \bibnamefont {Phillips}},\ and\
  \bibinfo {author} {\bibfnamefont {J.~V.}\ \bibnamefont {Porto}},\ }\bibfield
  {title} {\bibinfo {title} {Quantum and classical dynamics of a bose-einstein
  condensate in a large-period optical lattice},\ }\href
  {https://doi.org/10.1103/PhysRevA.80.043609} {\bibfield  {journal} {\bibinfo
  {journal} {Phys. Rev. A}\ }\textbf {\bibinfo {volume} {80}},\ \bibinfo
  {pages} {043609} (\bibinfo {year} {2009})}\BibitemShut {NoStop}%
\bibitem [{\citenamefont {Chalker}\ and\ \citenamefont
  {Shapiro}(2009)}]{Chalker:2009}%
  \BibitemOpen
  \bibfield  {author} {\bibinfo {author} {\bibfnamefont {J.~T.}\ \bibnamefont
  {Chalker}}\ and\ \bibinfo {author} {\bibfnamefont {B.}~\bibnamefont
  {Shapiro}},\ }\bibfield  {title} {\bibinfo {title} {Caustic formation in
  expanding condensates of cold atoms},\ }\href
  {https://doi.org/10.1103/PhysRevA.80.013603} {\bibfield  {journal} {\bibinfo
  {journal} {Phys. Rev. A}\ }\textbf {\bibinfo {volume} {80}},\ \bibinfo
  {pages} {013603} (\bibinfo {year} {2009})}\BibitemShut {NoStop}%
\bibitem [{\citenamefont {Mewes}\ \emph {et~al.}(1997)\citenamefont {Mewes},
  \citenamefont {Andrews}, \citenamefont {Kurn}, \citenamefont {Durfee},
  \citenamefont {Townsend},\ and\ \citenamefont {Ketterle}}]{Mewes1997}%
  \BibitemOpen
  \bibfield  {author} {\bibinfo {author} {\bibfnamefont {M.-O.}\ \bibnamefont
  {Mewes}}, \bibinfo {author} {\bibfnamefont {M.~R.}\ \bibnamefont {Andrews}},
  \bibinfo {author} {\bibfnamefont {D.~M.}\ \bibnamefont {Kurn}}, \bibinfo
  {author} {\bibfnamefont {D.~S.}\ \bibnamefont {Durfee}}, \bibinfo {author}
  {\bibfnamefont {C.~G.}\ \bibnamefont {Townsend}},\ and\ \bibinfo {author}
  {\bibfnamefont {W.}~\bibnamefont {Ketterle}},\ }\bibfield  {title} {\bibinfo
  {title} {{O}utput {C}oupler for {B}ose-{E}instein {C}ondensed {A}toms},\
  }\href {https://doi.org/10.1103/physrevlett.78.582} {\bibfield  {journal}
  {\bibinfo  {journal} {Phys. Rev. Lett.}\ }\textbf {\bibinfo {volume} {78}},\
  \bibinfo {pages} {582} (\bibinfo {year} {1997})}\BibitemShut {NoStop}%
\bibitem [{\citenamefont {Naraschewski}\ \emph {et~al.}(1997)\citenamefont
  {Naraschewski}, \citenamefont {Schenzle},\ and\ \citenamefont
  {Wallis}}]{Naraschewski1997}%
  \BibitemOpen
  \bibfield  {author} {\bibinfo {author} {\bibfnamefont {M.}~\bibnamefont
  {Naraschewski}}, \bibinfo {author} {\bibfnamefont {A.}~\bibnamefont
  {Schenzle}},\ and\ \bibinfo {author} {\bibfnamefont {H.}~\bibnamefont
  {Wallis}},\ }\bibfield  {title} {\bibinfo {title} {Phase diffusion and the
  output properties of a cw atom-laser},\ }\href
  {https://doi.org/10.1103/physreva.56.603} {\bibfield  {journal} {\bibinfo
  {journal} {Phys. Rev. A}\ }\textbf {\bibinfo {volume} {56}},\ \bibinfo
  {pages} {603} (\bibinfo {year} {1997})}\BibitemShut {NoStop}%
\bibitem [{\citenamefont {Ketterle}\ and\ \citenamefont
  {Miesner}(1997)}]{Ketterle:1997}%
  \BibitemOpen
  \bibfield  {author} {\bibinfo {author} {\bibfnamefont {W.}~\bibnamefont
  {Ketterle}}\ and\ \bibinfo {author} {\bibfnamefont {H.-J.}\ \bibnamefont
  {Miesner}},\ }\bibfield  {title} {\bibinfo {title} {Coherence properties of
  {Bose}-{Einstein} condensates and atom lasers},\ }\href@noop {} {\bibfield
  {journal} {\bibinfo  {journal} {Phys. Rev. A}\ }\textbf {\bibinfo {volume}
  {56}},\ \bibinfo {pages} {3291} (\bibinfo {year} {1997})}\BibitemShut
  {NoStop}%
\bibitem [{\citenamefont {Steck}\ \emph {et~al.}(1998)\citenamefont {Steck},
  \citenamefont {Naraschewski},\ and\ \citenamefont {Wallis}}]{Steck1998}%
  \BibitemOpen
  \bibfield  {author} {\bibinfo {author} {\bibfnamefont {H.}~\bibnamefont
  {Steck}}, \bibinfo {author} {\bibfnamefont {M.}~\bibnamefont
  {Naraschewski}},\ and\ \bibinfo {author} {\bibfnamefont {H.}~\bibnamefont
  {Wallis}},\ }\bibfield  {title} {\bibinfo {title} {Output of a pulsed atom
  laser},\ }\href {https://doi.org/10.1103/physrevlett.80.1} {\bibfield
  {journal} {\bibinfo  {journal} {Phys. Rev. Lett.}\ }\textbf {\bibinfo
  {volume} {80}},\ \bibinfo {pages} {1} (\bibinfo {year} {1998})}\BibitemShut
  {NoStop}%
\bibitem [{\citenamefont {Bloch}\ \emph {et~al.}(1999)\citenamefont {Bloch},
  \citenamefont {H{\"a}nsch},\ and\ \citenamefont {Esslinger}}]{Bloch1999}%
  \BibitemOpen
  \bibfield  {author} {\bibinfo {author} {\bibfnamefont {I.}~\bibnamefont
  {Bloch}}, \bibinfo {author} {\bibfnamefont {T.~W.}\ \bibnamefont
  {H{\"a}nsch}},\ and\ \bibinfo {author} {\bibfnamefont {T.}~\bibnamefont
  {Esslinger}},\ }\bibfield  {title} {\bibinfo {title} {Atom laser with a cw
  output coupler},\ }\href {https://doi.org/10.1103/physrevlett.82.3008}
  {\bibfield  {journal} {\bibinfo  {journal} {Phys. Rev. Lett.}\ }\textbf
  {\bibinfo {volume} {82}},\ \bibinfo {pages} {3008 } (\bibinfo {year}
  {1999})}\BibitemShut {NoStop}%
\bibitem [{\citenamefont {Schneider}\ and\ \citenamefont
  {Schenzle}(1999{\natexlab{a}})}]{Schneider1999}%
  \BibitemOpen
  \bibfield  {author} {\bibinfo {author} {\bibfnamefont {J.}~\bibnamefont
  {Schneider}}\ and\ \bibinfo {author} {\bibfnamefont {A.}~\bibnamefont
  {Schenzle}},\ }\bibfield  {title} {\bibinfo {title} {Output from an atom
  laser: theory vs. experiment},\ }\href
  {https://doi.org/10.1007/s003400050819} {\bibfield  {journal} {\bibinfo
  {journal} {Appl. Phys. B}\ }\textbf {\bibinfo {volume} {69}},\ \bibinfo
  {pages} {353} (\bibinfo {year} {1999}{\natexlab{a}})}\BibitemShut {NoStop}%
\bibitem [{\citenamefont {Ballagh}\ and\ \citenamefont
  {Savage}(2000)}]{Ballagh:2000}%
  \BibitemOpen
  \bibfield  {author} {\bibinfo {author} {\bibfnamefont {R.~J.}\ \bibnamefont
  {Ballagh}}\ and\ \bibinfo {author} {\bibfnamefont {C.~M.}\ \bibnamefont
  {Savage}},\ }\bibfield  {title} {\bibinfo {title} {The theory of atom
  lasers},\ }\href {https://doi.org/10.1142/9789812791900_0005} {\bibfield
  {journal} {\bibinfo  {journal} {Mod. Phys. Lett.}\ }\textbf {\bibinfo
  {volume} {B14}},\ \bibinfo {pages} {153} (\bibinfo {year}
  {2000})}\BibitemShut {NoStop}%
\bibitem [{\citenamefont {Bloch}\ \emph {et~al.}(2000)\citenamefont {Bloch},
  \citenamefont {H{\"a}nsch},\ and\ \citenamefont {Esslinger}}]{Bloch:2000}%
  \BibitemOpen
  \bibfield  {author} {\bibinfo {author} {\bibfnamefont {I.}~\bibnamefont
  {Bloch}}, \bibinfo {author} {\bibfnamefont {T.~W.}\ \bibnamefont
  {H{\"a}nsch}},\ and\ \bibinfo {author} {\bibfnamefont {T.}~\bibnamefont
  {Esslinger}},\ }\bibfield  {title} {\bibinfo {title} {Measurement of the
  spatial coherence of a trapped {Bose} gas at the phase transition},\ }\href
  {https://doi.org/10.1038/35003132} {\bibfield  {journal} {\bibinfo  {journal}
  {Nature}\ }\textbf {\bibinfo {volume} {403}},\ \bibinfo {pages} {166}
  (\bibinfo {year} {2000})}\BibitemShut {NoStop}%
\bibitem [{\citenamefont {Coq}\ \emph {et~al.}(2001)\citenamefont {Coq},
  \citenamefont {Thywissen}, \citenamefont {Rangwala}, \citenamefont {Gerbier},
  \citenamefont {Richard}, \citenamefont {Delannoy}, \citenamefont {Bouyer},\
  and\ \citenamefont {Aspect}}]{LeCoq2001}%
  \BibitemOpen
  \bibfield  {author} {\bibinfo {author} {\bibfnamefont {Y.~L.}\ \bibnamefont
  {Coq}}, \bibinfo {author} {\bibfnamefont {J.~H.}\ \bibnamefont {Thywissen}},
  \bibinfo {author} {\bibfnamefont {S.~A.}\ \bibnamefont {Rangwala}}, \bibinfo
  {author} {\bibfnamefont {F.}~\bibnamefont {Gerbier}}, \bibinfo {author}
  {\bibfnamefont {S.}~\bibnamefont {Richard}}, \bibinfo {author} {\bibfnamefont
  {G.}~\bibnamefont {Delannoy}}, \bibinfo {author} {\bibfnamefont
  {P.}~\bibnamefont {Bouyer}},\ and\ \bibinfo {author} {\bibfnamefont
  {A.}~\bibnamefont {Aspect}},\ }\bibfield  {title} {\bibinfo {title} {Atom
  laser divergence},\ }\href {https://doi.org/10.1103/physrevlett.87.170403}
  {\bibfield  {journal} {\bibinfo  {journal} {Phys. Rev. Lett.}\ }\textbf
  {\bibinfo {volume} {87}},\ \bibinfo {pages} {170403} (\bibinfo {year}
  {2001})}\BibitemShut {NoStop}%
\bibitem [{\citenamefont {Bloch}\ \emph {et~al.}(2001)\citenamefont {Bloch},
  \citenamefont {{K\"ohl}}, \citenamefont {Greiner}, \citenamefont
  {{H\"ansch}},\ and\ \citenamefont {Esslinger}}]{Bloch:2001}%
  \BibitemOpen
  \bibfield  {author} {\bibinfo {author} {\bibfnamefont {I.}~\bibnamefont
  {Bloch}}, \bibinfo {author} {\bibfnamefont {M.}~\bibnamefont {{K\"ohl}}},
  \bibinfo {author} {\bibfnamefont {M.}~\bibnamefont {Greiner}}, \bibinfo
  {author} {\bibfnamefont {T.~W.}\ \bibnamefont {{H\"ansch}}},\ and\ \bibinfo
  {author} {\bibfnamefont {T.}~\bibnamefont {Esslinger}},\ }\bibfield  {title}
  {\bibinfo {title} {Optics with an atom laser beam},\ }\href
  {https://doi.org/10.1103/physrevlett.87.030401} {\bibfield  {journal}
  {\bibinfo  {journal} {Phys. Rev. Lett.}\ }\textbf {\bibinfo {volume} {87}},\
  \bibinfo {pages} {030401} (\bibinfo {year} {2001})}\BibitemShut {NoStop}%
\bibitem [{\citenamefont {Chikkatur}\ \emph {et~al.}(2002)\citenamefont
  {Chikkatur}, \citenamefont {Shin}, \citenamefont {Leanhardt}, \citenamefont
  {Kielpinski}, \citenamefont {Tsikata}, \citenamefont {Gustavson},
  \citenamefont {Pritchard},\ and\ \citenamefont {Ketterle}}]{Chikkatur2002}%
  \BibitemOpen
  \bibfield  {author} {\bibinfo {author} {\bibfnamefont {A.~P.}\ \bibnamefont
  {Chikkatur}}, \bibinfo {author} {\bibfnamefont {Y.}~\bibnamefont {Shin}},
  \bibinfo {author} {\bibfnamefont {A.~E.}\ \bibnamefont {Leanhardt}}, \bibinfo
  {author} {\bibfnamefont {D.}~\bibnamefont {Kielpinski}}, \bibinfo {author}
  {\bibfnamefont {E.}~\bibnamefont {Tsikata}}, \bibinfo {author} {\bibfnamefont
  {T.~L.}\ \bibnamefont {Gustavson}}, \bibinfo {author} {\bibfnamefont {D.~E.}\
  \bibnamefont {Pritchard}},\ and\ \bibinfo {author} {\bibfnamefont
  {W.}~\bibnamefont {Ketterle}},\ }\bibfield  {title} {\bibinfo {title} {A
  continuous source of {Bose}-{Einstein} condensed atoms},\ }\href
  {https://doi.org/10.1126/science.296.5576.2193} {\bibfield  {journal}
  {\bibinfo  {journal} {Science}\ }\textbf {\bibinfo {volume} {296}},\ \bibinfo
  {pages} {2193} (\bibinfo {year} {2002})}\BibitemShut {NoStop}%
\bibitem [{\citenamefont {Haine}\ \emph {et~al.}(2002)\citenamefont {Haine},
  \citenamefont {Hope}, \citenamefont {Robins},\ and\ \citenamefont
  {Savage}}]{Haine:2002}%
  \BibitemOpen
  \bibfield  {author} {\bibinfo {author} {\bibfnamefont {S.~A.}\ \bibnamefont
  {Haine}}, \bibinfo {author} {\bibfnamefont {J.~J.}\ \bibnamefont {Hope}},
  \bibinfo {author} {\bibfnamefont {N.~P.}\ \bibnamefont {Robins}},\ and\
  \bibinfo {author} {\bibfnamefont {C.~M.}\ \bibnamefont {Savage}},\ }\bibfield
   {title} {\bibinfo {title} {Stability of continuously pumped atom lasers},\
  }\href {https://doi.org/10.1103/physrevlett.88.170403} {\bibfield  {journal}
  {\bibinfo  {journal} {Phys. Rev. Lett.}\ }\textbf {\bibinfo {volume} {88}},\
  \bibinfo {pages} {170403} (\bibinfo {year} {2002})}\BibitemShut {NoStop}%
\bibitem [{\citenamefont {Lee}\ \emph {et~al.}(2015)\citenamefont {Lee},
  \citenamefont {Haine}, \citenamefont {Bradley},\ and\ \citenamefont
  {Davis}}]{Lee:2015}%
  \BibitemOpen
  \bibfield  {author} {\bibinfo {author} {\bibfnamefont {G.~M.}\ \bibnamefont
  {Lee}}, \bibinfo {author} {\bibfnamefont {S.~A.}\ \bibnamefont {Haine}},
  \bibinfo {author} {\bibfnamefont {A.~S.}\ \bibnamefont {Bradley}},\ and\
  \bibinfo {author} {\bibfnamefont {M.~J.}\ \bibnamefont {Davis}},\ }\bibfield
  {title} {\bibinfo {title} {Coherence and linewidth of a continuously pumped
  atom laser at finite temperature},\ }\href
  {https://doi.org/10.1103/physreva.92.013605} {\bibfield  {journal} {\bibinfo
  {journal} {Phys. Rev. A}\ }\textbf {\bibinfo {volume} {92}},\ \bibinfo
  {pages} {013605} (\bibinfo {year} {2015})}\BibitemShut {NoStop}%
\bibitem [{\citenamefont {Harvie}\ \emph {et~al.}(2020)\citenamefont {Harvie},
  \citenamefont {Butcher},\ and\ \citenamefont {Goldwin}}]{Harvie:2020}%
  \BibitemOpen
  \bibfield  {author} {\bibinfo {author} {\bibfnamefont {G.}~\bibnamefont
  {Harvie}}, \bibinfo {author} {\bibfnamefont {A.}~\bibnamefont {Butcher}},\
  and\ \bibinfo {author} {\bibfnamefont {J.}~\bibnamefont {Goldwin}},\
  }\bibfield  {title} {\bibinfo {title} {Coherence time of a cold-atom laser
  below threshold},\ }\href {https://doi.org/10.1364/ol.402975} {\bibfield
  {journal} {\bibinfo  {journal} {Optics Letters}\ }\textbf {\bibinfo {volume}
  {45}},\ \bibinfo {pages} {5448} (\bibinfo {year} {2020})}\BibitemShut
  {NoStop}%
\bibitem [{\citenamefont {Riou}\ \emph {et~al.}(2008)\citenamefont {Riou},
  \citenamefont {Coq}, \citenamefont {Impens}, \citenamefont {Guerin},
  \citenamefont {Bord{\'{e}}}, \citenamefont {Aspect},\ and\ \citenamefont
  {Bouyer}}]{Riou:2008}%
  \BibitemOpen
  \bibfield  {author} {\bibinfo {author} {\bibfnamefont {J.-F.}\ \bibnamefont
  {Riou}}, \bibinfo {author} {\bibfnamefont {Y.~L.}\ \bibnamefont {Coq}},
  \bibinfo {author} {\bibfnamefont {F.}~\bibnamefont {Impens}}, \bibinfo
  {author} {\bibfnamefont {W.}~\bibnamefont {Guerin}}, \bibinfo {author}
  {\bibfnamefont {C.~J.}\ \bibnamefont {Bord{\'{e}}}}, \bibinfo {author}
  {\bibfnamefont {A.}~\bibnamefont {Aspect}},\ and\ \bibinfo {author}
  {\bibfnamefont {P.}~\bibnamefont {Bouyer}},\ }\bibfield  {title} {\bibinfo
  {title} {Theoretical tools for atom-laser-beam propagation},\ }\href
  {https://doi.org/10.1103/PhysRevA.77.033630} {\bibfield  {journal} {\bibinfo
  {journal} {Physical Review A}\ }\textbf {\bibinfo {volume} {77}},\ \bibinfo
  {pages} {033630} (\bibinfo {year} {2008})}\BibitemShut {NoStop}%
\bibitem [{\citenamefont {Busch}\ \emph {et~al.}(2002)\citenamefont {Busch},
  \citenamefont {Köhl}, \citenamefont {Esslinger},\ and\ \citenamefont
  {M{\o}lmer}}]{Busch:2002}%
  \BibitemOpen
  \bibfield  {author} {\bibinfo {author} {\bibfnamefont {T.}~\bibnamefont
  {Busch}}, \bibinfo {author} {\bibfnamefont {M.}~\bibnamefont {Köhl}},
  \bibinfo {author} {\bibfnamefont {T.}~\bibnamefont {Esslinger}},\ and\
  \bibinfo {author} {\bibfnamefont {K.}~\bibnamefont {M{\o}lmer}},\ }\bibfield
  {title} {\bibinfo {title} {Transverse mode of an atom laser},\ }\href
  {https://doi.org/10.1103/PhysRevA.65.043615} {\bibfield  {journal} {\bibinfo
  {journal} {Physical Review A}\ }\textbf {\bibinfo {volume} {65}},\ \bibinfo
  {pages} {043615} (\bibinfo {year} {2002})}\BibitemShut {NoStop}%
\bibitem [{\citenamefont {Köhl}\ \emph {et~al.}(2005)\citenamefont {Köhl},
  \citenamefont {Busch}, \citenamefont {M{\o}lmer}, \citenamefont {Hänsch},\
  and\ \citenamefont {Esslinger}}]{Kohl:2005}%
  \BibitemOpen
  \bibfield  {author} {\bibinfo {author} {\bibfnamefont {M.}~\bibnamefont
  {Köhl}}, \bibinfo {author} {\bibfnamefont {T.}~\bibnamefont {Busch}},
  \bibinfo {author} {\bibfnamefont {K.}~\bibnamefont {M{\o}lmer}}, \bibinfo
  {author} {\bibfnamefont {T.~W.}\ \bibnamefont {Hänsch}},\ and\ \bibinfo
  {author} {\bibfnamefont {T.}~\bibnamefont {Esslinger}},\ }\bibfield  {title}
  {\bibinfo {title} {Observing the profile of an atom laser beam},\ }\href
  {https://doi.org/10.1103/PhysRevA.72.063618} {\bibfield  {journal} {\bibinfo
  {journal} {Physical Review A}\ }\textbf {\bibinfo {volume} {72}},\ \bibinfo
  {pages} {063618} (\bibinfo {year} {2005})}\BibitemShut {NoStop}%
\bibitem [{\citenamefont {Riou}\ \emph {et~al.}(2006)\citenamefont {Riou},
  \citenamefont {Guerin}, \citenamefont {Coq}, \citenamefont {Fauquembergue},
  \citenamefont {Josse}, \citenamefont {Bouyer},\ and\ \citenamefont
  {Aspect}}]{Riou:2006}%
  \BibitemOpen
  \bibfield  {author} {\bibinfo {author} {\bibfnamefont {J.-F.}\ \bibnamefont
  {Riou}}, \bibinfo {author} {\bibfnamefont {W.}~\bibnamefont {Guerin}},
  \bibinfo {author} {\bibfnamefont {Y.~L.}\ \bibnamefont {Coq}}, \bibinfo
  {author} {\bibfnamefont {M.}~\bibnamefont {Fauquembergue}}, \bibinfo {author}
  {\bibfnamefont {V.}~\bibnamefont {Josse}}, \bibinfo {author} {\bibfnamefont
  {P.}~\bibnamefont {Bouyer}},\ and\ \bibinfo {author} {\bibfnamefont
  {A.}~\bibnamefont {Aspect}},\ }\bibfield  {title} {\bibinfo {title} {{Beam
  Quality of a Nonideal Atom Laser}},\ }\href
  {https://doi.org/10.1103/PhysRevLett.96.070404} {\bibfield  {journal}
  {\bibinfo  {journal} {Physical Review Letters}\ }\textbf {\bibinfo {volume}
  {96}},\ \bibinfo {pages} {070404} (\bibinfo {year} {2006})}\BibitemShut
  {NoStop}%
\bibitem [{\citenamefont {Dall}\ \emph {et~al.}(2007)\citenamefont {Dall},
  \citenamefont {Byron}, \citenamefont {Truscott}, \citenamefont {Dennis},
  \citenamefont {Johnsson}, \citenamefont {Jeppesen},\ and\ \citenamefont
  {Hope}}]{Dall:2007}%
  \BibitemOpen
  \bibfield  {author} {\bibinfo {author} {\bibfnamefont {R.~G.}\ \bibnamefont
  {Dall}}, \bibinfo {author} {\bibfnamefont {L.~J.}\ \bibnamefont {Byron}},
  \bibinfo {author} {\bibfnamefont {A.~G.}\ \bibnamefont {Truscott}}, \bibinfo
  {author} {\bibfnamefont {G.~R.}\ \bibnamefont {Dennis}}, \bibinfo {author}
  {\bibfnamefont {M.~T.}\ \bibnamefont {Johnsson}}, \bibinfo {author}
  {\bibfnamefont {M.}~\bibnamefont {Jeppesen}},\ and\ \bibinfo {author}
  {\bibfnamefont {J.~J.}\ \bibnamefont {Hope}},\ }\bibfield  {title} {\bibinfo
  {title} {Observation of transverse interference fringes on an atom laser
  beam},\ }\href {https://doi.org/10.1364/OE.15.017673} {\bibfield  {journal}
  {\bibinfo  {journal} {Optics Express}\ }\textbf {\bibinfo {volume} {15}},\
  \bibinfo {pages} {17673} (\bibinfo {year} {2007})}\BibitemShut {NoStop}%
\bibitem [{\citenamefont {DeWitt-Morette}\ \emph {et~al.}(1983)\citenamefont
  {DeWitt-Morette}, \citenamefont {Nelson},\ and\ \citenamefont
  {Zhang}}]{DeWitt-Morette:1983}%
  \BibitemOpen
  \bibfield  {author} {\bibinfo {author} {\bibfnamefont {C.}~\bibnamefont
  {DeWitt-Morette}}, \bibinfo {author} {\bibfnamefont {B.}~\bibnamefont
  {Nelson}},\ and\ \bibinfo {author} {\bibfnamefont {T.-R.}\ \bibnamefont
  {Zhang}},\ }\bibfield  {title} {\bibinfo {title} {Caustic problems in quantum
  mechanics with applications to scattering theory},\ }\href
  {https://doi.org/10.1103/physrevd.28.2526} {\bibfield  {journal} {\bibinfo
  {journal} {Phys. Rev. D}\ }\textbf {\bibinfo {volume} {28}},\ \bibinfo
  {pages} {2526} (\bibinfo {year} {1983})}\BibitemShut {NoStop}%
\bibitem [{\citenamefont {Cartier}\ and\ \citenamefont
  {DeWitt-Morette}(2006)}]{Cartier:2006}%
  \BibitemOpen
  \bibfield  {author} {\bibinfo {author} {\bibfnamefont {P.}~\bibnamefont
  {Cartier}}\ and\ \bibinfo {author} {\bibfnamefont {C.}~\bibnamefont
  {DeWitt-Morette}},\ }\href {https://doi.org/10.1017/cbo9780511535062} {\emph
  {\bibinfo {title} {Functional Integration: Action and Symmetries}}},\
  Cambridge Monographs on Mathematical Physics\ (\bibinfo  {publisher}
  {Cambridge University Press},\ \bibinfo {year} {2006})\BibitemShut {NoStop}%
\bibitem [{\citenamefont {Whitney}(1955)}]{Whitney:1955}%
  \BibitemOpen
  \bibfield  {author} {\bibinfo {author} {\bibfnamefont {H.}~\bibnamefont
  {Whitney}},\ }\bibfield  {title} {\bibinfo {title} {On singularities of
  mappings of euclidean spaces. i. mappings of the plane into the plane},\
  }\href {https://doi.org/10.1007/978-1-4612-2972-8_27} {\bibfield  {journal}
  {\bibinfo  {journal} {The Annals of Mathematics}\ }\textbf {\bibinfo {volume}
  {62}},\ \bibinfo {pages} {374} (\bibinfo {year} {1955})}\BibitemShut
  {NoStop}%
\bibitem [{\citenamefont {Dyke}(1982)}]{VanDyke:1982}%
  \BibitemOpen
  \bibfield  {author} {\bibinfo {author} {\bibfnamefont {M.~V.}\ \bibnamefont
  {Dyke}},\ }\href {https://doi.org/10.1115/1.3167071} {\emph {\bibinfo {title}
  {A Album of Fluid Motion}}},\ \bibinfo {edition} {14th}\ ed.\ (\bibinfo
  {publisher} {Parabolic Press, Inc.},\ \bibinfo {year} {1982})\BibitemShut
  {NoStop}%
\bibitem [{\citenamefont {Samimy}\ \emph {et~al.}(2004)\citenamefont {Samimy},
  \citenamefont {Breuer}, \citenamefont {Leal},\ and\ \citenamefont
  {Steen}}]{Samimy:2004}%
  \BibitemOpen
  \bibinfo {editor} {\bibfnamefont {M.}~\bibnamefont {Samimy}}, \bibinfo
  {editor} {\bibfnamefont {K.}~\bibnamefont {Breuer}}, \bibinfo {editor}
  {\bibfnamefont {L.}~\bibnamefont {Leal}},\ and\ \bibinfo {editor}
  {\bibfnamefont {P.}~\bibnamefont {Steen}},\ eds.,\ \href
  {https://doi.org/10.1017/CBO9780511610820} {\emph {\bibinfo {title} {A
  Gallery of Fluid Motion}}}\ (\bibinfo  {publisher} {Cambridge University
  Press},\ \bibinfo {year} {2004})\BibitemShut {NoStop}%
\bibitem [{\citenamefont {Chandrasekhar}(1943)}]{Chandrasekhar:1943}%
  \BibitemOpen
  \bibfield  {author} {\bibinfo {author} {\bibfnamefont {S.}~\bibnamefont
  {Chandrasekhar}},\ }\href@noop {} {\emph {\bibinfo {title} {On the decay of
  plane shock waves.}}},\ \bibinfo {type} {Tech. Rep.}\ \bibinfo {number}
  {423}\ (\bibinfo  {institution} {Ballistic Research Laboratories},\ \bibinfo
  {address} {Aberdeen Proving Ground, Md.},\ \bibinfo {year}
  {1943})\BibitemShut {NoStop}%
\bibitem [{\citenamefont {Henson}\ \emph {et~al.}(2018)\citenamefont {Henson},
  \citenamefont {Yue}, \citenamefont {Hodgman}, \citenamefont {Shin},
  \citenamefont {Smirnov}, \citenamefont {Ostrovskaya}, \citenamefont {Guan},\
  and\ \citenamefont {Truscott}}]{Henson:2018}%
  \BibitemOpen
  \bibfield  {author} {\bibinfo {author} {\bibfnamefont {B.~M.}\ \bibnamefont
  {Henson}}, \bibinfo {author} {\bibfnamefont {X.}~\bibnamefont {Yue}},
  \bibinfo {author} {\bibfnamefont {S.~S.}\ \bibnamefont {Hodgman}}, \bibinfo
  {author} {\bibfnamefont {D.~K.}\ \bibnamefont {Shin}}, \bibinfo {author}
  {\bibfnamefont {L.~A.}\ \bibnamefont {Smirnov}}, \bibinfo {author}
  {\bibfnamefont {E.~A.}\ \bibnamefont {Ostrovskaya}}, \bibinfo {author}
  {\bibfnamefont {X.~W.}\ \bibnamefont {Guan}},\ and\ \bibinfo {author}
  {\bibfnamefont {A.~G.}\ \bibnamefont {Truscott}},\ }\bibfield  {title}
  {\bibinfo {title} {{Bogoliubov}-{Cherenkov} radiation in an atom laser},\
  }\href {https://doi.org/10.1103/physreva.97.063601} {\bibfield  {journal}
  {\bibinfo  {journal} {Phys. Rev. A}\ }\textbf {\bibinfo {volume} {97}},\
  \bibinfo {pages} {063601} (\bibinfo {year} {2018})}\BibitemShut {NoStop}%
\bibitem [{\citenamefont {DeWitt-Morette}\ and\ \citenamefont
  {Cartier}(1997)}]{DeWitt-Morette:1997}%
  \BibitemOpen
  \bibfield  {author} {\bibinfo {author} {\bibfnamefont {C.}~\bibnamefont
  {DeWitt-Morette}}\ and\ \bibinfo {author} {\bibfnamefont {P.}~\bibnamefont
  {Cartier}},\ }\bibinfo {title} {Physics on and near caustics}\ (\bibinfo
  {publisher} {Springer US},\ \bibinfo {address} {Boston, MA},\ \bibinfo {year}
  {1997})\ pp.\ \bibinfo {pages} {51--66}\BibitemShut {NoStop}%
\bibitem [{\citenamefont {Adhikari}\ and\ \citenamefont
  {Hussein}(2008)}]{Adhikari:2008b}%
  \BibitemOpen
  \bibfield  {author} {\bibinfo {author} {\bibfnamefont {S.~K.}\ \bibnamefont
  {Adhikari}}\ and\ \bibinfo {author} {\bibfnamefont {M.~S.}\ \bibnamefont
  {Hussein}},\ }\bibfield  {title} {\bibinfo {title} {Semiclassical scattering
  in two dimensions},\ }\href {https://doi.org/10.1119/1.2970054} {\bibfield
  {journal} {\bibinfo  {journal} {Amer. J. Phys.}\ }\textbf {\bibinfo {volume}
  {76}},\ \bibinfo {pages} {1108} (\bibinfo {year} {2008})}\BibitemShut
  {NoStop}%
\bibitem [{\citenamefont {Goussev}\ and\ \citenamefont
  {Richter}(2013)}]{Goussev:2013}%
  \BibitemOpen
  \bibfield  {author} {\bibinfo {author} {\bibfnamefont {A.}~\bibnamefont
  {Goussev}}\ and\ \bibinfo {author} {\bibfnamefont {K.}~\bibnamefont
  {Richter}},\ }\bibfield  {title} {\bibinfo {title} {Scattering of quantum
  wave packets by shallow potential islands: {A} quantum lens},\ }\href
  {https://doi.org/10.1103/PhysRevE.87.052918} {\bibfield  {journal} {\bibinfo
  {journal} {Phys. Rev. E}\ }\textbf {\bibinfo {volume} {87}},\ \bibinfo
  {pages} {052918} (\bibinfo {year} {2013})}\BibitemShut {NoStop}%
\bibitem [{\citenamefont {Edwards}\ \emph {et~al.}(1999)\citenamefont
  {Edwards}, \citenamefont {Griggs}, \citenamefont {Holman}, \citenamefont
  {Clark}, \citenamefont {Rolston},\ and\ \citenamefont
  {Phillips}}]{Edwards:1999}%
  \BibitemOpen
  \bibfield  {author} {\bibinfo {author} {\bibfnamefont {M.}~\bibnamefont
  {Edwards}}, \bibinfo {author} {\bibfnamefont {D.~A.}\ \bibnamefont {Griggs}},
  \bibinfo {author} {\bibfnamefont {P.~L.}\ \bibnamefont {Holman}}, \bibinfo
  {author} {\bibfnamefont {C.~W.}\ \bibnamefont {Clark}}, \bibinfo {author}
  {\bibfnamefont {S.~L.}\ \bibnamefont {Rolston}},\ and\ \bibinfo {author}
  {\bibfnamefont {W.~D.}\ \bibnamefont {Phillips}},\ }\bibfield  {title}
  {\bibinfo {title} {Properties of a raman atom-laser output coupler},\ }\href
  {https://doi.org/10.1088/0953-4075/32/12/312} {\bibfield  {journal} {\bibinfo
   {journal} {J. Phys. B}\ }\textbf {\bibinfo {volume} {32}},\ \bibinfo {pages}
  {2935} (\bibinfo {year} {1999})}\BibitemShut {NoStop}%
\bibitem [{\citenamefont {Schneider}\ and\ \citenamefont
  {Schenzle}(1999{\natexlab{b}})}]{Schneider:1999}%
  \BibitemOpen
  \bibfield  {author} {\bibinfo {author} {\bibfnamefont {J.}~\bibnamefont
  {Schneider}}\ and\ \bibinfo {author} {\bibfnamefont {A.}~\bibnamefont
  {Schenzle}},\ }\bibfield  {title} {\bibinfo {title} {Output from an atom
  laser: theory vs. experiment},\ }\href
  {https://doi.org/10.1007/s003400050819} {\bibfield  {journal} {\bibinfo
  {journal} {Applied Physics B}\ }\textbf {\bibinfo {volume} {69}},\ \bibinfo
  {pages} {353} (\bibinfo {year} {1999}{\natexlab{b}})}\BibitemShut {NoStop}%
\bibitem [{\citenamefont {Schneider}\ and\ \citenamefont
  {Schenzle}(2000)}]{Schneider:2000}%
  \BibitemOpen
  \bibfield  {author} {\bibinfo {author} {\bibfnamefont {J.}~\bibnamefont
  {Schneider}}\ and\ \bibinfo {author} {\bibfnamefont {A.}~\bibnamefont
  {Schenzle}},\ }\bibfield  {title} {\bibinfo {title} {Investigations of a
  two-mode atom-laser model},\ }\href
  {https://doi.org/10.1103/physreva.61.053611} {\bibfield  {journal} {\bibinfo
  {journal} {Phys. Rev. A}\ }\textbf {\bibinfo {volume} {61}},\ \bibinfo
  {pages} {053611} (\bibinfo {year} {2000})}\BibitemShut {NoStop}%
\bibitem [{\citenamefont {Heller}(1984)}]{Heller:1984}%
  \BibitemOpen
  \bibfield  {author} {\bibinfo {author} {\bibfnamefont {E.~J.}\ \bibnamefont
  {Heller}},\ }\bibfield  {title} {\bibinfo {title} {Bound-state eigenfunctions
  of classically chaotic hamiltonian systems: Scars of periodic orbits},\
  }\href {https://doi.org/10.1103/PhysRevLett.53.1515} {\bibfield  {journal}
  {\bibinfo  {journal} {Phys. Rev. Lett.}\ }\textbf {\bibinfo {volume} {53}},\
  \bibinfo {pages} {1515} (\bibinfo {year} {1984})}\BibitemShut {NoStop}%
\bibitem [{\citenamefont {Arora}\ \emph {et~al.}(2011)\citenamefont {Arora},
  \citenamefont {Safronova},\ and\ \citenamefont {Clark}}]{Arora:2011}%
  \BibitemOpen
  \bibfield  {author} {\bibinfo {author} {\bibfnamefont {B.}~\bibnamefont
  {Arora}}, \bibinfo {author} {\bibfnamefont {M.~S.}\ \bibnamefont
  {Safronova}},\ and\ \bibinfo {author} {\bibfnamefont {C.~W.}\ \bibnamefont
  {Clark}},\ }\bibfield  {title} {\bibinfo {title} {Tune-out wavelengths of
  alkali-metal atoms and their applications},\ }\href
  {https://doi.org/10.1103/PhysRevA.84.043401} {\bibfield  {journal} {\bibinfo
  {journal} {Phys. Rev. A}\ }\textbf {\bibinfo {volume} {84}},\ \bibinfo
  {pages} {043401} (\bibinfo {year} {2011})}\BibitemShut {NoStop}%
\bibitem [{\citenamefont {Forbes}(2021)}]{gitlab:catastrophe_atom_optics}%
  \BibitemOpen
  \bibfield  {author} {\bibinfo {author} {\bibfnamefont {M.~M.}\ \bibnamefont
  {Forbes}},\ }\href@noop {} {\bibinfo {title} {Code and {3\textsc{d}}
  visualizations of the caustic surfaces}},\ \bibinfo {howpublished}
  {\href{https://gitlab.com/coldatoms/publications/catastrophe_atom_optics}{gitlab.com/coldatoms/publications/catastrophe\_atom\_optics}}
  (\bibinfo {year} {2021})\BibitemShut {NoStop}%
\bibitem [{\citenamefont {Mossman}\ \emph {et~al.}(2021)\citenamefont
  {Mossman}, \citenamefont {Bersano}, \citenamefont {Forbes},\ and\
  \citenamefont {Engels}}]{osf:catastrophe_atom_optics}%
  \BibitemOpen
  \bibfield  {author} {\bibinfo {author} {\bibfnamefont {M.~E.}\ \bibnamefont
  {Mossman}}, \bibinfo {author} {\bibfnamefont {T.~M.}\ \bibnamefont
  {Bersano}}, \bibinfo {author} {\bibfnamefont {M.~M.}\ \bibnamefont
  {Forbes}},\ and\ \bibinfo {author} {\bibfnamefont {P.}~\bibnamefont
  {Engels}},\ }\href {https://doi.org/10.17605/OSF.IO/KDM9S} {\bibinfo {title}
  {Catastrophe atom optics: Data and figures}},\ \bibinfo {howpublished}
  {(\href{https://osf.io/kdm9s}{osf.io/kdm9s})} (\bibinfo {year}
  {2021})\BibitemShut {NoStop}%
\end{thebibliography}
\fi

\begin{acknowledgments}
We thank Mark Hoefer for helpful initial discussions and D.~H.~J.~O'Dell for discussions relating caustics and chaos.
T.~M.~B.\@, M.~E.~M.\@ and P.~E.\@ are supported by the
\gls{NSF} through Grants No. \textsc{phy-1912540}.
P.~E. acknowledges support through the Ralph G. Yount Distinguished Professorship at WSU.
M.~E.~M acknowledges support through the Clare Boothe Luce Program of the Henry Luce Foundation.
M.~M.~F.\@ is supported by the \gls{NSF} through Grant No.\@ \textsc{phy-1707691}.
\end{acknowledgments}

\section{Author Contributions}
M.E.M. and P.E. conceived the experiment.
M.E.M., T.M.B. and P.E. performed experiments and data analysis.
M.M.F. performed theoretical calculations and numerical simulations.
All authors discussed the results and contributed to the writing of the manuscript.

\section{Competing Interests}
The authors declare no competing interests.

\section{Materials \& Correspondence}
Please direct any questions or requests concerning this article to P.~Engels or M.~M.~Forbes.

\end{document}